  \documentclass[12pt,doublespace, onecolumn]{IEEEtran}
\usepackage{amsmath,epsfig}
\usepackage{graphics,amsmath,amsfonts,amssymb,epsfig,latexsym,amsfonts,graphicx,amssymb}
\usepackage{subfigure}   
\usepackage{cite}
\usepackage{bm}
\usepackage{makecell}
\usepackage{color}
\usepackage{diagbox} 
\usepackage{stfloats}  
\usepackage{footnote}
\makesavenoteenv{table}  
\newtheorem{theorem}{Theorem}
\newtheorem{lemma}{Lemma} 
\newtheorem{definition}[theorem]{Definition}
\newtheorem{proposition}{Proposition}  

\newtheorem{coro}{Corollary}
\newtheorem{rmk}{Remark}
\newtheorem{claim}{Claim}

\title{ Joint Downlink Base Station Association and Power Control for Max-Min Fairness: Computation and Complexity }
\author{Ruoyu Sun\thanks{.....}}

\author{Ruoyu Sun, Mingyi Hong, Zhi-Quan Luo\thanks{This  research is
supported in part by a research gift from Huawei Technologies Inc. Part of this paper is published in SPAWC 2012 \cite{Sun2012}.}}

\begin{document}
\newcommand{\tr}{{\rm Tr}}
\newcommand{\st}{{\rm s.t.}}
\newcommand{\rank}{{\rm rank}}
\maketitle
\begin{abstract}
In a heterogeneous network (HetNet) with a large number of low power base stations (BSs),
 proper user-BS association and power control is crucial to achieving desirable system performance.
In this paper, we systematically study the joint BS association and power allocation problem for
a downlink cellular network under the max-min fairness criterion.
 First, we show that this problem is NP-hard.
 Second, we show that the upper bound of the optimal value can be easily computed, and propose a two-stage algorithm to find a high-quality suboptimal solution.
Simulation results show that the proposed algorithm is near-optimal in the high-SNR regime.
 Third, we show that the problem under some additional mild assumptions can be solved to global optima in polynomial time by a semi-distributed algorithm. This result is based on a transformation of the original problem to an assignment problem with gains $\log(g_{ij})$, where $\{g_{ij}\}$ are the channel gains.

\end{abstract}
\section{Introduction}
\label{sec:intro}

Wireless cellular networks are increasingly relying on using low power transmit nodes such as pico base stations (BS) to provide substantially improved data service. It is predicted that the number of such low power BSs will grow by an order of magnitude in the next few years, and it may soon exceed the number of mobile users \cite{Andrews13}.
Such densely deployed heterogeneous network (HetNet) architecture has made the user-BS association a key design issue. The conventional greedy scheme that associates receivers with the transmitter providing the strongest signal is no longer effective in a HetNet, as such scheme may cause severe load imbalance. A variant of the greedy scheme called Range Extension \cite{LTEHetNet} can provide better load balance in the downlink by re-adjusting cell boundaries, but is still highly suboptimal during periods of congestion.

A systematic approach for the user-BS association problem is to jointly design the association and power allocation so as to maximize a network-wide utility. Early works in this direction \cite{Yates95,Hanly1995} proposed a fixed point iteration to jointly adjust BS association and power allocation  in the uplink. The goal was to minimize the total transmit power subject to certain QoS (Quality of Service) constraints for each user.
It was shown in \cite{Yates95,Hanly1995} that this algorithm converges to a globally optimal solution, provided that the QoS constraints are feasible. This algorithm has later been extended to accommodate extra power budget constraints \cite{Yates1997}, as well as to work for the downlink setting \cite{RFLiuDL}.
The fixed point algorithm in \cite{Yates95,Hanly1995} can also be interpreted as an alternating optimization approach: fix the BS association, each user updates its power to satisfy the SINR (Signal to Interference plus Noise Ratio) constraint; fix the power, each user updates its BS association to maximize its SINR. This alternating optimization approach was extended to joint beamforming, power allocation and BS association in the uplink \cite{RFLiu98}.

 Recently, there is a surge of renewed interests in the BS association problem, mainly due to the key role it plays  in the resource allocation for the HetNet \cite{Sardellitti13, Madan2010, Kuang12, Hong2012}. Alternating optimization is again a frequently used approach, in which a system utility maximization problem is solved by alternately optimizing over BS association and other system parameters.
   For instances, reference \cite{Madan2010} proposed to maximize the utility by iteratively updating
   resource partitioning, power allocation and BS association. Reference \cite{Kuang12} proposed to maximize the sum rate by iteratively updating power allocation and BS association. However, since the association variable is discrete, it is not known whether these algorithms converge to local optima.
 A game theoretic approach was adopted in \cite{Hong2012}, where the BS association problem is formulated as a noncooperative game and solved by a best-response type algorithm with guaranteed convergence to a Nash equilibrium of the game (typically not a local optima of the sum utility).

To circumvent the difficulty of dealing with discrete variables, researchers have proposed to eliminate the discrete variables either by introducing
power/beamformer variables for each user-BS pair \cite{Smolyar09,Sparse2013} or by relaxing the binary BS association variables  to continuous variables in $[0,1]$ \cite{Ye2012,Sun2013}.
Either approach relies on the CoMP (Coordinated Multiple Point) transmission strategy, i.e. one user can be served by multiple BSs that share data via backhaul links. If the system does not support CoMP transmission, the multiple-BS association needs to be converted to a single-BS association such as in \cite{Ye2012}, resulting in suboptimal solutions.

The computational complexity of maximizing a certain utility function by joint BS association and power allocation has been studied in different scenarios \cite{Hong2012,Maz2012,SunUL}.
For the sum rate utility function, the NP hardness of the joint design problem has been established for both the uplink transmission \cite{Hong2012}
and the downlink transmission \cite{Maz2012}.
For the {\it uplink} max-min fairness problem, reference \cite{SunUL} proposed a fixed point algorithm that converges to the global optima in a geometric rate, 
{\color{black} which implies the pseudo-polynomial time solvability of this problem (note that pseudo-polynomial time solvability does not imply polynomial time solvability). 
It is not known whether there exists a polynomial time algorithm for the same problem in either the {\it uplink} or the {\it downlink} direction.  }


In this paper, we systematically investigate the max-min fair downlink joint BS association and power allocation problem, under the per-BS power constraint (referred to as problem $(\rm P)$).
Our main contributions are summarized as follows:
\begin{itemize}
\item [(1)] We prove that problem $(\rm P)$ is NP-hard (Theorem \ref{theoremNP}).
Note that with fixed BS association, achieving downlink max-min fairness by using power allocation is polynomial time solvable since it can be solved by a binary search where each subproblem is a linear program \cite{Luo2008}.
Our results show that the joint design problem is intrinsically much more difficult than the power allocation problem.


\item [(2)]
 We propose a fixed point algorithm \emph{ULSum} to compute the upper bound on the optimal value of (P) by solving a relaxed version of (P).
Based on this fixed point algorithm, we then propose a two-stage algorithm \emph{DLSumA} to find a suboptimal solution to problem (P).
Our simulations show that \emph{DLSumA} achieves an objective value that is close to the upper bound in many cases.


{\color{black}
\item [(3)]
 We prove that when there are an equal number of BSs and users, problem $(\rm P)$ under additional constraints that the SINR of each user is at least $0$ dB
 is polynomial time solvable (Theorem \ref{main result}).
 The problem can be solved to global optima by a two-stage semi-distributed algorithm \emph{AUFP}: in stage 1, compute the BS association by using the auction algorithm \cite{Bert92} to solve a maximum weighted bipartite matching problem with weights $\{\log(g_{ij})\}$; in stage 2, use a fixed point iteration to solve the power allocation problem.
 Notably, the auction algorithm in the first stage can be implemented in a fully distributed manner.
}
\end{itemize}

We summarize the computational
 complexity results related to our problem in Table \ref{table of summary of complexity}.
Note that all the results are for the case with per-BS power constraint.

\begin{table*}[!htbp]\label{table of summary of complexity}\normalsize
\caption{  \small Summary of the Complexity Status of the Joint BS Association and Power Control Problem }\footnotesize
\begin{tabular}{|c|c|c|c|c|c|}
\hline 
\diaghead{\theadfont Problem Class Objective}{Objective}{Setting} & \thead{Fixed BS association} & \thead{UL (Uplink) joint} & \thead{DL (Downlink) joint} & \thead{ {\color{black}UL or DL} joint, equal number \\ of BSs and users, $\mathrm{SINR}_k \geq 1, \forall k$}\\  
\hline
\thead{Maximize Sum-rate} & NP-hard \cite{Luo2008} & NP-hard \cite{Hong2012} & NP-hard \cite{Maz2012} & Unknown \\
\hline
\thead{Maximize Min-rate} & Polynomial \cite{Luo2008} & {\color{black} Pseudo-polynomial \cite{SunUL} }
& NP-hard (Theorem \ref{theoremNP}) & Polynomial (Theorem \ref{main result}) \\
\hline
\end{tabular}
\end{table*}


Remark: The third part mentioned above has appeared in a previous conference publication \cite{Sun2012} with the difference that \cite{Sun2012} presents the results and algorithms for uplink transmission. Nevertheless, for one-to-one matching of users and BSs, the uplink and downlink problems are essentially the same.  

The rest of the paper is organized as follows.
In Section \ref{sec: System Model}, we describe the system model and provide the complexity result.
 {In Section \ref{sec: SumPower and Algorithms}, we first consider the sum power constrained problem and propose a fixed point algorithm to solve it to global
 optima.
 We then describe a two-stage fixed point algorithm to solve $(\rm P)$ and discuss two techniques to improve it. }
 In Section \ref{sec: one-to-one matching}, we show the polynomial-time solvability of $(\rm P_1)$ with additional SINR assumptions,
 and present a semi-distributed algorithm to solve $(\rm P_1)$ to global optima.
Numerical experiments are reported in Section~\ref{sec:simulation} to evaluate the performance of the proposed algorithms. %

\section{System Model, Complexity Analysis and Algorithm for a Subproblem}
\label{sec: System Model}
Consider a downlink wireless network where $N$ base stations (BS) intend to transmit data to $K$ mobile users. Both the BSs and the users are equipped with a single antenna, and they share the same time/frequency resource for transmission. Each user is to be associated to exactly one BS, but one BS can serve multiple users. Our goal is to maximize the minimum rate by joint BS association and power allocation, subject to the power constraint of each BS.

Denote by $g_{nk}$ the channel gain between user $k$ and BS $n$.
Let $\bm a= (a_1,a_2,\dots, a_K)$ denote the association profile, i.e.,
$a_k = n$ if user $k$ is associated with BS $n$. 
Let $\Omega_n$ denote the serving set of BS $n$, i.e.
\begin{equation}\label{serving set def}
 \Omega_n \triangleq \{k \mid a_k = n \}.
 \end{equation}
Denote by $p_k$ the power BS $a_k$ used to transmit the data intended for user $k$.
Suppose the power budget of BS $n$ is $\bar{p}_n$, then $ \sum_{k \in \Omega_n} p_k \leq \bar{p}_n $.
Let $\bar{\bm p} = (\bar{p}_1, \dots, \bar{p}_N)$, and $\bm{p}=(p_1,\cdots,p_K)$.

The problem of maximizing the minimum SINR by joint downlink BS association and power allocation is formulated as follows:
\begin{equation}\label{max min}
\begin{split}
({\rm P}): \max_{\mbox{$\bm{p}$},\mbox{$\bm{a}$}  } &  \min_{k=1,\dots,K}   {\rm SINR}_k \triangleq \frac{ p_{k} g_{a_k k} }{\sigma_{k}^2 + \sum_{i \neq k}p_{i} g_{a_i k}  }, \\
\st \;      & \quad  p_k \geq 0, \;  k=1,\dots,K,   \\
            & \quad \sum_{k \in \Omega_n } p_k \leq \bar{p}_n, \;  n=1,\dots,N, \\
            & \quad  a_k \in \{1,2,\dots, N \},  \; k=1,\dots,K, 
\end{split}
\end{equation}
where $\sigma_{k}^2$ is the receive noise power at user $k$.


Optimizing over $\bm a$ with a fixed $\bm p$ seems to be difficult. One possible reason is that the power constraints $\sum_{k \in \Omega_n } p_k \leq \bar{p}_n$
depend on $\bm a$ since the serving sets $\Omega_n$'s are defined by $\bm a$.
One can expect that solving problem (\ref{max min}) over both $\bm p$ and $\bm a$ is also difficult.
Indeed, Theorem \ref{theoremNP} below shows that the max-min fairness problem by joint BS association and power control with per-BS power constraint is NP-hard.
The proof of this result is given in Appendix \ref{appen: proof of NP hard}.

\begin{theorem}\label{theoremNP}
{\it Problem $(\rm P)$, i.e. finding the optimal BS association and
power control that maximize the minimum SINR, is NP-hard
in general.}
\end{theorem}

Optimizing over $\bm p$ with fixed $\bm a$ is easy.
 Given a BS association $\bm a$, problem (\ref{max min}) is a max-min fairness power control problem for an IBC (Interfering Broadcast Channel). It can be solved in polynomial time using a binary search strategy whereby a QoS constrained subproblem is solved by an LP (Linear Programming) at each step \cite{Luo2008}. In practice, the problem can be alternatively solved to global optima by a fixed point algorithm, which is presented next.

\subsection{A Fixed Point Algorithm for Power Allocation in IBC}\label{subsec: fixed BS, update power}

In this subsection, we propose a fixed point algorithm to solve the power control problem with fixed BS association.
Suppose the BS association $\bm a $ is fixed, we want to solve the following max-min fairness power control problem for an IBC:
\begin{equation}\label{max min2}
\begin{split}
(\mathrm P_{\bm a}): \max_{\mbox{$\bm p$}   } &  \min_{k=1,\dots,K}   {\rm SINR}_k \triangleq \frac{ p_{k} g_{a_k k} }{\sigma_{k}^2 + \sum_{i \neq k}p_{i} g_{a_i k}  }, \\
\st         & \quad \quad  p_k \geq 0, \quad  k=1,\dots,K,   \\
            & \quad \sum_{k \in \Omega_n } p_k \leq \bar{p}_n, \quad  n=1,\dots,N,
\end{split}
\end{equation}
where $\Omega_n = \{k \mid a_k = n \}$ is fixed. 

As mentioned before, problem $(\mathrm P_{\bm a})$ can be solved by a binary search strategy whereby a QoS constrained subproblem is solved by an LP at each step. Since the binary search can be time-consuming, we present a fixed point algorithm that directly solves $(\mathrm P_{\bm a})$ (see \cite{SunUL} for a comparison of the binary search method and the fixed point algorithm for the uplink max-min fairness problem by joint BS association and power control).
This algorithm is a generalization of the algorithm in \cite{TanChiang2009} for SISO IC (Interference Channel) where one BS serves exactly one user.

Define
\begin{equation}\label{def of M}
\begin{split}
 M_k(\bm p) & \triangleq   \frac{   \sigma_{k}^2 + \sum_{i \neq k} p_i g_{a_i k}  }{ g_{a_k k} },   \\
 M(\bm p)         & \triangleq (M_1(\bm p), \dots, M_K(\bm p)),
\end{split}
\end{equation}
 Notice that $M_k(\bm p)$ represents the minimum power needed by BS $a_k$ to achieve an SINR value of $1$ with fixed $p_j, \;\forall\;j\neq k$.
Define 
\begin{equation}\label{Omega norm def}
\begin{split}
\Omega &  \triangleq \{ \Omega_1, \dots, \Omega_N \}=\{1,\cdots,K\},   \\
\| \bm p \|_{\Omega} &  \triangleq \max_n \frac{ (\sum_{k \in \Omega_n } p_k)}{ \bar{p}_n }.
\end{split}
\end{equation}
The power constraints of $(\mathrm P_{\bm a})$ can be rewritten as
$$ \| \bm p \|_{\Omega} \leq 1. $$

The proposed algorithm picks a random positive power vector $\bm p(0)$, and updates the power vector as follows:  
\begin{equation}\label{IBC fixed point}
 \bm p(t+1) \leftarrow  \frac{ M( \bm p(t)) }{ \| M( \bm p(t)) \|_{\Omega} }.
 \end{equation}

Consider a special case where $K = N, \Omega_n = \{n \}$ and $\bar{p}_n = P_{\rm max}$, i.e. BS $n$ only serves user $n$ and all BSs have the same power budget.
Then the proposed fixed point algorithm \eqref{IBC fixed point} becomes
$$
p_k(t+1) \leftarrow  \frac{ M_k( \bm p(t)) }{ \max_j M_j( \bm p(t)) } P_{\rm max},
$$
which is exactly \cite[Algorithm 3]{TanChiang2009}.

In \cite[Algorithm 3]{TanChiang2009}, $\|\cdot \|_{\Omega}$ becomes a weighted $\ell_{\infty}$-norm, while in our algorithm,
$\|\cdot \|_{\Omega}$ is a weighted $\ell_{\infty}$/$\ell_{1}$-norm.
By an argument similar to \cite[Theorem 2]{SunUL}, we can prove that \eqref{IBC fixed point} converges, at a geometric rate, to the optimal solution of $(\mathrm P_{\bm a})$, which is also the solution to the following fixed point equation:
$$
\bm p = \frac{ M( \bm p ) }{ \| M( \bm p ) \|_{\Omega} } .
$$
We omit the proof due to space reason.

\section{ An Upper Bound and A Two-stage Fixed-point Algorithm   }
\label{sec: SumPower and Algorithms}
In this section, we first compute an upper bound of the optimal value of (P), by solving a relaxed version of (P).
We then propose a two-stage algorithm to find a high-quality suboptimal solution to problem (P): the first stage determines the association by solving the relaxed version of (P), and the second stage computes the optimal power allocation corresponding to the association obtained earlier.

\subsection{An Upper Bound via the Sum Power Relaxation}\label{upper bound}
As mentioned in Section \ref{sec: System Model}, one difficulty in solving (P) is that the power constraints $\sum_{k \in \Omega_n } p_k \leq \bar{p}_n,\;\forall\; n$
involve the discrete variable $\bm a$.
To circumvent this difficulty, one can replace the per-BS power constraint in problem \eqref{max min} by a sum power constraint to obtain the following problem:
\begin{equation}\label{max min_sum power}
\begin{split}
(\rm P_{\rm sum}): \max_{\mbox{$\bm p$},\mbox{$\bm a$}  } &  \min_{k=1,\dots,K}   {\rm SINR}_k \triangleq \frac{ p_{k} g_{a_k k} }{\sigma_{k}^2 + \sum_{i \neq k}p_{i} g_{a_i k}  }, \\
\st \;      & \quad  p_k \geq 0, \;  k=1,\dots,K,   \\
            & \quad \sum_{k=1 }^K p_k \leq \| \bar{\bm p} \|_1, \\
            & \quad  a_k \in \{1,2,\dots, N \},  \; k=1,\dots,K, 
\end{split}
\end{equation}
where $\|\bar{\bm p}\|_1 = \sum_{n=1}^N \bar{p}_n$.

Problem $(\rm P_{\rm sum})$ is a relaxation of problem (P) because any $(\bm p, \bm a)$ satisfying the individual power constraints of problem (P)
also satisfies the sum power constraints of $(\rm P_{\rm sum})$. Therefore, the optimal value of $(\rm P_{\rm sum})$, which is an upper bound of the optimal value of (P), can be used to benchmark algorithms that directly solve (P). Note that formulation $(\rm P_{\rm sum})$ itself is not as interesting as the original problem since the BSs typically cannot share transmit power. The problem $(\rm P_{\rm sum})$ with fixed $\bm a$ can also be interpreted as a power control problem in the broadcast channel where there are a single BS with $N$ antennas and $K$ users (see, e.g., \cite{Boche04Duality} for a similar formulation).

The benefit of replacing the per BS power constraint by the sum power constraint is that we can utilize the uplink-downlink duality {\color{black} (see, e.g., \cite[Theorem 1]{Boche04Duality}).}
Consider the uplink problem
\begin{equation}\label{max min, sum power, UL}
\begin{split}
(\rm P_{\rm sum}^{UL}): \max_{\mbox{$\bm p$},\mbox{$\bm a$}  } &  \min_{k=1,\dots,K}   {\rm SINR}_k \triangleq \frac{  g_{a_k k} p_{ k} }{\delta_{a_k}^2 + \sum_{j \neq k} g_{a_k j } p_{ j} }, \\
\st \;      & \quad  p_k \geq 0, \;  k=1,\dots,K,   \\
            & \quad \sum_{k=1 }^K p_k \leq \|\bar{\bm p}\|_1, \\
            & \quad  a_k \in \{1,2,\dots, N \},  \; k=1,\dots,K, 
\end{split}
\end{equation}
where $\delta_n^2$ is the noise power at BS $n$.
According to \cite[Theorem 1]{Boche04Duality}, with fixed BS association $\bm a$, if $\delta_n^2 = \sigma_k^2 = \sigma^2, \;\forall\; k,n$, then the uplink problem $(\rm P_{\rm sum}^{UL})$ and the downlink problem
$(\rm P_{\rm sum})$ have the same optimal value. Therefore, the original problems $(\rm P_{\rm sum}^{UL})$ and $(\rm P_{\rm sum})$ (with $(\bm p,\bm a)$ being the variable) also have the same optimal value and the same optimal BS association. Such an uplink-downlink duality result is formally stated as below.
\begin{proposition}\label{prop 2}  \it
 Suppose $(\bm p^{\rm UL}, \bm a^{\rm UL})$ is an optimal solution of $(\rm P_{\rm sum}^{UL})$.
If $\delta_n^2 = \sigma_k^2 = \sigma^2, \;\forall\; k,n$ (i.e. equal noise power), then the following two uplink-downlink duality properties hold: 
\begin{itemize}
\item [(1)] $\gamma^{UL}$, the min-SINR achieved by  $(\bm p^{\rm UL}, \bm a^{\rm UL})$, is also the optimal value of the downlink problem $(\rm P_{\rm sum})$;
\item [(2)] $\bm a^{\rm UL}$ is also the optimal BS association for the downlink problem $(\rm P_{\rm sum})$.
\end{itemize}
\end{proposition}
Note that $\bm p^{\rm UL}$ is not the optimal downlink power allocation for $(\rm P_{\rm sum})$ in general.

To solve $(\rm P_{\rm sum}^{UL})$, we propose a fixed point algorithm which is similar to the algorithm in \cite{SunUL}.
 Define
\begin{align}
 T_k^{(n)}(\bm p) & \triangleq \left\{ \frac{   \sigma_{n}^2 + \sum_{j \neq k} g_{n j} p_j }{ g_{n k} }  \right\} ,  \label{define of Tkn}     \\
 T_k(\bm p)       & \triangleq \min_{ 1\leq n\leq N} T_k^{(n)}(\bm p), \label{def of T_k} \\
 T(\bm p)         & \triangleq (T_1(\bm p), \dots, T_K(\bm p)),  \label{def of T}  \\
  A_k (\bm p)     & \triangleq \arg \min_{ n } T_k^{(n)}(\bm p).  \label{def of optimal a}
\end{align}

Notice that $T_k^{(n)}(\bm p)$ represents the minimum amount of power needed by user $k$ to achieve an SINR value of $1$, if its associated BS is $n$ and the power of other users are fixed at $p_j, \forall j\neq k$.
The minimum power user $k$ needs to achieve an SINR level of $1$ among all possible choices of BS association is defined as $ T_k(\bm p)$, and the corresponding
BS association is defined as $A_k (\bm p)$ (if there are multiple elements in $\arg \min_n T_k^{(n)}(\bm p)$, let $A_k (\bm p)$ be any one of them). The proposed algorithm is summarized in Table \ref{table of NFP for UL sum}.
\begin{table}[htbp]
\caption{\normalsize \emph{ULSum}: A fixed point algorithm for $(\rm P_{\rm sum}^{UL})$ } \normalsize
\begin{tabular}{p{460pt}}
\hline \\
{\bf Initialization}: pick random positive power vector $\bm p(0)$.
\\
Loop $t$:
\\ 1) Compute BS association: $a_k(t) \leftarrow A_k(\bm p(t)), \ \forall\; k $.
\\ 2) Update power: $ \bm p(t+1) \leftarrow  \frac{ T( \bm p(t)) }{ \| T( \bm p(t)) \|_1 } \sum_n \bar{p}_n   $ ;
\\ $\quad \quad $ where $ \| \bm p\|_1 = \sum_k p_k $.
\\ {\bf Iterate until}: $\|\bm p(t) - \bm p(t+1) \| \leq \epsilon \|\bm \bar{\bm{p}} \| $ for some $\epsilon>0$. 
\\
\\
\hline
\end{tabular}\label{table of NFP for UL sum}
\end{table}
By similar arguments as in \cite[Theorem 2]{SunUL} and \cite[Theorem 4]{Hanly1995}, we can prove the following convergence result (see a detailed proof in Appendix \ref{appen: proof of UL_DL duality}).
\begin{proposition}\label{propULconverge}
\it
Suppose the noise power $\delta_n^2>0, \forall n$.
Then the uplink problem $(\rm P_{\rm sum}^{UL})$ has a unique optimal power vector $\bm p^{\rm UL}$ and the algorithm ULSum generates a sequence $\bm p(t)$ that converges to $\bm p^{\rm UL}$ at a geometric rate.
Denote $\mathcal{A}^{\rm UL}$ as the set of optimal BS associations for the problem $(\rm P_{\rm sum}^{UL})$,
then the sequence $\bm a(t)$ generated by ULSum satisfies that there exists $T$ such that $\bm a(t) \in \mathcal{A}^{\rm UL}$ for all $t \geq T$.
\end{proposition}

Proposition \ref{propULconverge} implies that if $(\rm P_{\rm sum}^{UL})$ has a unique optimal BS association, then \emph{ULSum} finds this BS association after finite iterations. Theoretically speaking, it is possible that $(\rm P_{\rm sum}^{UL})$ has more than one optimal BS associations, but
in the numerical experiments with random channel gains we find that $(\rm P_{\rm sum}^{UL})$ always has a unique optimal BS association.

\subsection{Two-stage Fixed Point Algorithm to Solve $(\mathrm P)$}\label{two tech for Algo 2}
%
%
 One may consider using the solution obtained by \emph{ULSum} as an approximate solution to $(\rm P)$. However this may be problematic because $\bm p^{\rm UL}$ is not always feasible for problem $(\rm P)$. Therefore a second stage is needed to find an admissible power allocation.  A natural way to do this is to solve a restricted version of $(\rm P)$ with fixed BS association $ \bm a^{\mathrm{UL} } \in \mathcal{A}^{\rm UL}$ by the algorithm proposed in section \ref{subsec: fixed BS, update power}.
The basic algorithm we propose for solving problem $(\mathrm P)$, named ``\emph{DLSum}'', simply combines \emph{ULSum} and the fixed-point iteration \eqref{IBC fixed point}. It is summarized in Table \ref{table of Algorithm 2}.
\begin{table}[htbp]
\caption{\emph{DLSum}: A Two-Stage Algorithm for $(\mathrm P)$ }\normalsize
\begin{tabular}{p{460pt}}
\hline \vspace{0.1cm}
Stage 1: Compute a BS association $\bm a^{UL}$ that is optimal for $(\mathrm P_{\rm sum})$ by \emph{ULSum}.
\\
Stage 2: Given $\bm a = \bm a^{UL}$, compute $\bm p^{\rm DL}$ by \eqref{IBC fixed point}.
\\ \\
\hline
\end{tabular}\label{table of Algorithm 2}
\end{table}

In the following, we extend algorithm \emph{DLSum} to improve its performance. We will discuss two techniques, both of which can better exploit the special characteristics of the HetNet. 
We use the notation $\mathrm P (g_{nk}, \bar{p}_n )$ to represent an instance of problem $(\mathrm P)$ with channel gains $\{ g_{nk} \}$ and power budgets $\{ \bar{p}_n \}$; we use  $\mathrm{val} \left( \mathrm P (g_{nk}, \bar{p}_n ) \right)$ to denote its optimal value.
It is easy to show that for any positive weights $\{\alpha_n\}$, we have
\begin{equation}\label{reweighting does not affect}
 \mathrm{val} \left( \mathrm P (g_{nk}, \bar{p}_n ) \right) = \mathrm{val} \left( \mathrm P (g_{nk}/\alpha_n, \alpha_n \bar{p}_n ) \right).
 \end{equation}

Similarly, we use $\mathrm P_{\mathrm{sum}} (g_{nk}, \sum_n \bar{p}_n )$ to denote an instance of problem $(\rm P_{sum})$ with channel gains $\{ g_{nk} \}$ and a sum power budget $ \sum_{n=1}^N \bar{p}_n $, and use $\mathrm{val} \left( \mathrm P_{\mathrm{sum}} (g_{nk}, \sum_n \bar{p}_n ) \right)$ to denote its optimal value.
In general, $ \mathrm{val} \left( \mathrm P_{\mathrm{sum}} (g_{nk}, \sum_n \bar{p}_n ) \right)$ may not be equal to $\mathrm{val} \left( \mathrm P_{\mathrm{sum}} (g_{nk}/\alpha_n, \sum_n \alpha_n \bar{p}_n ) \right).$
As mentioned in Section \ref{upper bound}, $\mathrm{val} \left( \mathrm P_{\mathrm{sum}} (g_{nk}/\alpha_n, \sum_n \alpha_n \bar{p}_n ) \right)$ is an upper bound of $ \mathrm{val} \left( \mathrm P (g_{nk}/\alpha_n, \alpha_n \bar{p}_n ) \right)$. Combining with \eqref{reweighting does not affect}, we obtain
\begin{equation}\label{upper bound holds for any reweights}
\mathrm{val} \left( \mathrm P (g_{nk}, \bar{p}_n ) \right)\leq  \mathrm{val} ( \mathrm P_{\mathrm{sum}} ( g_{nk}/\alpha_n, \sum_n \alpha_n \bar{p}_n ) ),\;
\forall\; \alpha_n >0.
\end{equation}
This says problem $\mathrm P_{\mathrm{sum}} (g_{nk}/\alpha_n, \sum_n \alpha_n \bar{p}_n )$ is also a relaxation of ${\rm P}(g_{nk}/\alpha_n, \alpha_n \bar{p}_n )$. The following discussion depends heavily on this key observation.


The first technique, called ``power balancing'', intends to eliminate the {\it power imbalance} effect, which arises in the HetNet when there is a huge difference in the transmit power available to different kinds of BSs.
The idea is to scale the power budgets and channel gains simultaneously, so that the original problem is transformed to the one that has the same power budget for each BS.
To this end, we choose the weights $\{\alpha_n\}$ to be inversely proportional to the power budgets $\{\bar{p}_n\}$, i.e. $$ \alpha_n = \bar{p}_{\rm max}/\bar{p}_n, $$
where $ \bar{p}_{\rm max} \triangleq \max_n \bar{p}_n$.
By doing the scaling, the original problem $\mathrm P (g_{nk}, \bar{p}_n)$ is transformed to a new problem $\mathrm P (\bar{p}_n g_{nk}/\bar{p}_{\rm max}, \bar{p}_{\rm max})$,
which has the same power budget $\bar{p}_{\rm max}$ for each BS.
We combine \emph{DLSum} (or \emph{ULSum}) with the technique ``power balancing'' as follows: replace $g_{nk}$ by $\bar{p}_n g_{nk}/\bar{p}_{\rm max}$ and replace $\bar{p}_n$ by $\bar{p}_{\rm max}$ for all $k,n$, then apply \emph{DLSum} (or \emph{ULSum}).
The modified algorithm \emph{ULSum} with ``power balancing'', referred to as \emph{ULSumA} (meaning \emph{ULSum-Advanced}), provides a new upper bound of the optimal value of $\rm(P)$ according to \eqref{upper bound holds for any reweights} (for a special choice of $\{\alpha_n\}$).



The second technique, called ``effective sum-power'', is based on the following observation: with fixed BS association $\bm a$, at the optimality of $(\mathrm P_{\bm a})$ often there are only a few BSs transmitting with full power, while the rest use a small portion of their individual power budget.
This observation implies that the total power consumed by the BSs is usually much less than the sum of the power budgets.
 Therefore, the relaxed sum power constraint $\sum_n p_n \leq \sum_n \bar{p}_n$ can be very loose, in which case $\mathrm{val} (\mathrm P_{\mathrm{sum}} (g_{nk}, \bar{p}_n ))$ also becomes a loose upper bound for $\mathrm{val} (\mathrm P (g_{nk}, \bar{p}_n ))$.
A tighter upper bound can be obtained as follows. Suppose the optimal solution of the original problem $\mathrm P (g_{nk}, \bar{p}_n )$ is
$(\bm p^*, \bm a^*)$. Obviously replacing the original power constraints $p_n \leq \bar{p}_n$ by the effective power constraints $p_n \leq p^*_n$ does not change the optimal value, i.e.
\begin{equation}\label{effective sum power}
\mathrm{val}(\mathrm P (g_{nk}, \bar{p}_n ))=\mathrm{val}(\mathrm P (g_{nk}, p^*_n )).
\end{equation}

Relaxing $\mathrm P (g_{nk}, p^*_n )$ to the sum power constrained problem $\mathrm P_{\mathrm{sum}} (g_{nk}, \sum_n p^*_n )$, and combining with \eqref{effective sum power}, we obtain a new upper bound of the original optimal value:
\begin{equation}\label{upper bound by true optimal power}
\mathrm{val} ( \mathrm P (g_{nk}, \bar{p}_n ) )\leq  \mathrm{val} ( \mathrm P_{\mathrm{sum}} (g_{nk}, \sum_n p^*_n ) ).
\end{equation}
The new sum power budget $\sum_n p^*_n$ is usually strictly less than $\sum_n \bar{p}_n$, in which case we obtain a strictly better upper bound: $$\mathrm{val} ( \mathrm P_{\mathrm{sum}} (g_{nk}, \sum_n p^*_n ) ) < \mathrm{val} ( \mathrm P_{\mathrm{sum}} (g_{nk}, \sum_n \bar{p}_n ) ).$$
 {\color{black}
 Of course, $\sum_n p^*_n$ is an unknown value if the optimal power vector $\bm p^*$ is unknown, thus in practice we can only find an approximation of $\sum_n p^*_n$.
For any BS association $\bm a$, denote $\bm p(\bm a) = (p_1(\bm a), \dots, p_N(\bm a))$ as the optimal power vector corresponding to $\bm a$.
We propose to approximate the unknown $\sum_n p^*_n = \sum_n p_n(\bm a^*)$ by $\sum_n p_n(\bm a^{\mathrm{UL}}) = \sum p_n^{\mathrm{DL}}$, where
 $\bm a^*$ is an optimal BS association and $\bm a^{\mathrm{UL}}$ is the BS association computed in \emph{ULSum} and can be viewed as an approximation of $\bm a^*$, and $\bm p^{\mathrm{DL}}$ is the power vector computed in Stage 2 of \emph{DLSum}.
Therefore, we apply the proposed technique ``effective sum-power'' to \emph{DLSum} as follows: first obtain
 $\bm p^{\mathrm{DL}}$ by \emph{DLSum}, then run \emph{DLSum} again with $\sum_n p^{\mathrm{DL}}_n$ being the sum power budget for Stage 1.
 } 

Combining the two techniques, we obtain an improved version of \emph{DLSum}, which is summarized in Table \ref{table of Algorithm 3, modified heuristic}.
Note that Step 0 represents the technique ``power balancing'', while Step 1, Step 2 and the new sum power constrained problem in Step 3 represent the technique ``effective sum-power''.
\begin{table}[htbp]\label{table of Algorithm 3, modified heuristic}
\caption{ \emph{DLSumA} (Advanced version of \emph{DLSum}) }\normalsize 
\begin{tabular}{p{460pt}}
\hline
\\
Step 0: Scaling: $g_{nk} \longleftarrow \bar{p}_n g_{nk}/\bar{p}_{\rm max}, \forall\ n,k$. \\
Step 1: Solve $\mathrm P^{\mathrm{UL}}_{\mathrm{sum}} (g_{nk}, N \bar{p}_{\rm max}) $ by \emph{ULSum}. \\
   $\quad \quad \quad $ Denote the computed BS association as $\bm a^{\rm UL}$. \\
Step 2: Given $\bm a = \bm a^{UL}$, compute $\bm p^{\rm DL}$ by \eqref{IBC fixed point}.  \\
Step 3: Solve $\mathrm P^{\mathrm{UL}}_{\mathrm{sum}} (g_{nk}, \sum_n p^{\mathrm{DL}}_n )$ by \emph{ULSum}. \\
   $\quad \quad \quad $ Denote the computed BS association as $ \hat{\bm a}^{\rm UL}$. \\
Step 4: Given $\hat{\bm a}^{\rm UL}$, compute $ \hat{\bm p}^{\rm DL}$ by \eqref{IBC fixed point}.  \\
\\
\hline
\end{tabular}
\end{table}

The combination of the two ideas of \eqref{upper bound holds for any reweights} and \eqref{upper bound by true optimal power} leads to the following upper bound of the original optimal value:
\begin{equation}\label{best upper bound}
\mathrm{val} \left( \mathrm P (g_{nk}, \bar{p}_n ) \right) \leq \min_{\alpha_n > 0, \forall n} \mathrm{val} ( \mathrm P_{\mathrm{sum}} (g_{nk}/\alpha_n, \sum_n \alpha_n p^*_n ) ).
\end{equation}
\emph{DLSumA} can be viewed as a method to compute an approximation of the right hand side of \eqref{best upper bound}: replacing the minimum
over all possible $\{ \alpha_n \}$ by a special choice of $\{\alpha_n \}$ that balances the transmit power, and replacing the unknown optimal sum power by
a special sum power.
\section{ One-to-One Matching: Optimality and Semi-Distributed Algorithm    } 
\label{sec: one-to-one matching}

In this section, we study a simplified version of the original problem $(\rm P)$, with two extra assumptions: \\
\begin{equation}\label{Assump A}
\begin{split}
\!\!\!\!\!\!\!\text{\emph{Assumption A}: } & \text{There are an equal number of }  \\
   & \text{BSs and users, i.e. } K =N.\quad\quad\quad\quad \
\end{split}
\end{equation}
\begin{equation}
\label{Assump B}
\!\text{\;\emph{Assumption B}: } \text{Each BS serves exactly one user. }\quad
\end{equation}

%
%

We remark that Assumption A-B are quite standard in the literature. For example the resulting problem appears in \cite{Madan2010,Ye2012,Rosenberg2013} as a subproblem to the joint design of BS association and scheduling. 
Under these assumptions, the BS association $\bm a$ becomes a one-to-one matching between BSs and users, and the max-min fairness problem $(\rm P)$ becomes the following problem:
\begin{equation}\label{max min, one-to-one}
\begin{split}
(\rm P_1): \max_{\mbox{$\bm p$},\mbox{$\bm a$}  } &  \min_{k=1,\dots,K}   {\rm SINR}_k \triangleq \frac{ p_{k} g_{a_k k} }{\sigma_{k}^2 + \sum_{i \neq k}p_{i} g_{a_i k}  }, \\
\st \;      & \quad  0 \leq p_k \leq \bar{p}_{a_k}, \;  k=1,\dots,K, \\
            & \quad  \bm a \text{ is a permutation of } \{1,2,\dots, K \}. 
\end{split}
\end{equation}


 It turns out that problem $(\rm P_1)$ is again NP-hard in general:
\begin{coro}\label{coro:NPhard of P1}
{\it Problem $(\rm P_1)$, i.e. finding the globally optimal one-to-one user-BS matching and power control that
maximize the minimum SINR, is NP hard in general.}
\end{coro}
 The proof of this result is similar to that of Theorem \ref{theoremNP}. In fact we only need to consider the first two configurations in the proof of Theorem \ref{theoremNP} (cf. Fig. \ref{figConfiguration}).







\subsection{Polynomial Time Solvability}
In this subsection, we consider problem (\ref{max min, one-to-one}) with additional QoS constraints:
\begin{equation}\label{max min, one-to-one,SINR>1}
\begin{split}
(\mathrm P_1^{\prime}): \max_{\bm p,\bm a  } &  \min_{k=1,\dots,K}   {\rm SINR}_k \triangleq \frac{ p_{k} g_{a_k k} }{\sigma_{k}^2 + \sum_{j \neq k}p_{j} g_{a_j k}  }, \\
\st \;      & \quad  0 \leq p_k \leq \bar{p}_{a_k}, \;  k=1,\dots,K, \\
            & \quad  \bm a \text{ is a permutation of } \{1,2,\dots, K \}, \\
            & \quad {\rm SINR}_k  \geq 1, k=1,\dots, K. 
\end{split}
\end{equation}
Note that we have added the constraints that the SINR of each user is at least $0$dB. Such constraints are reasonable in practice, because they merely require that the received signal power should be no less than the interference plus noise power.

 Assumption B (the assumption of one-to-one matching) can be relaxed with the additional QoS constraints: in fact, $(\mathrm P_1^{\prime})$ is equivalent to
the following problem (see Theorem \ref{main result}):
\begin{equation}\label{max min, one-to-one,SINR>1}
\begin{split}
(\mathrm P^{\prime}): \max_{\bm p,\bm a  } &  \min_{k=1,\dots,K}   {\rm SINR}_k \triangleq \frac{ p_{k} g_{a_k k} }{\sigma_{k}^2 + \sum_{j \neq k}p_{j} g_{a_j k}  }, \\
\st \;      & \quad  0 \leq p_k \leq \bar{p}_{a_k}, \;  k=1,\dots,K, \\
            & \quad  \bm a_k \in \{1,2,\dots, K \}, k=1,\dots, K, \\
            & \quad {\rm SINR}_k  \geq 1, k=1,\dots, K. 
\end{split}
\end{equation}

 The following result shows that adding the QoS constraints makes problem $(\mathrm P_1)$ and $(\rm P)$ (when $K=N$) tractable.
\begin{theorem}\label{main result}
{\it Problems $(\mathrm P_1^{\prime})$ and $(\mathrm P^{\prime})$, the QoS constrained joint BS association and power control problems, are equivalent and they are polynomial time solvable. }
\end{theorem}

{\color{black}
Remark: The reference \cite[Theorem 2]{Sun2012} has shown that a similar uplink problem is polynomial time solvable. For one-to-one matchings of BSs and users, the uplink and downlink problems are essentially the same.
}

\textbf{Proof of Theorem \ref{main result}}
We first prove that problem $(\mathrm P_1^{\prime})$ is polynomial time solvable.
We say a BS association $\bm{a} = (a_1,\dots,a_K)$ is a feasible BS association for problem $(\mathrm P_1^{\prime})$ if there is a power vector $\textbf{p}$ such that
$$
{\rm SINR}_k  = \frac{ p_{k} g_{a_k k} }{\sigma_{k}^2 + \sum_{j \neq k}p_{j} g_{a_j k}  } \geq 1, \;k=1,\dots, K.
$$

The key observation, which will be explained shortly, is that there is at most one feasible BS association for problem $(\mathrm P_1^{\prime})$. In addition, the unique candidate for a feasible BS association is the solution of a maximum weighted matching problem (also called assignment problem), which is known to be polynomial time solvable.  

To facilitate discussion, we need the following definition.
\begin{definition}{(Assignment problem)}
Consider $K$ persons and $K$ objects, where $G_{ik}$ is the gain of assigning object $i$ to person $k$.
The assignment problem is to find a one-to-one assignment $\bm{a} = (a_1,\dots,a_K)$,
  such that the total gain $\sum_{k=1}^K G_{a_k k} $ is maximized.
\end{definition}



The following key lemma builds the connection between the assignment problem and the BS association problem.
The proof is relegated to Appendix \ref{appen: proof of assignment lemma}.
\begin{lemma}\label{key lemma}
If $\bm{a}^*$ is feasible for problem $(\mathrm P_1^{\prime})$, then $\bm{a}^*$ is the unique optimal solution to the assignment problem
with gains $\{\log(g_{ij})\}$.
\end{lemma}

\begin{figure}[htbp]
    \vspace{-.1cm}
    \centering
{\includegraphics[width=0.65\linewidth]{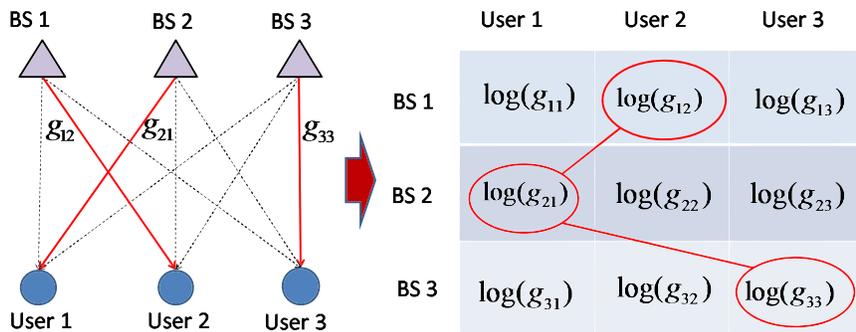} }
\vspace{-.5cm}
  \caption{ Relation between the BS association problem and the assignment problem for a $3 \times 3$ network:
  if $\bm a^* = (2,1,3)$ is feasible for $(\mathrm P_1^{\prime})$, then $(2,1,3)$ is the solution to the assignment problem with gains $\{\log g_{ij}\}$,
  i.e. $\log(g_{21}) + \log(g_{12})+ \log(g_{33})$ is the maximal total gain among all gains of the form $\sum_{i=1}^3 \log g_{a_i i}$.
}\label{Fig1matching} \vspace{-.1cm}
\end{figure}

We illustrate Lemma \ref{key lemma} through a simple example in Fig. \ref{Fig1matching}.

After computing the unique candidate $\bm{a}^*$, we only need to solve problem $(\mathrm P_1^{\prime})$ with fixed BS association $\bm{a} = \bm{a}^*$. We present a centralized algorithm that solves problem $(\mathrm P_1^{\prime})$ in Table \ref{Hungarian}.
\begin{table}[htbp]
\caption{Polynomial Time Algorithm for Problem $(\mathrm P_1^{\prime})$ }
\label{Hungarian}
\normalsize
\begin{tabular}{p{460pt}}
\hline
\\
\textbf{Stage 1}: Solve the assignment problem with gains $\{\log(g_{ij})\}$
using Hungarian algorithm \cite{HungarianAlgo}.  Denote the optimal solution as $ \hat{\bm a}$.\\
\textbf{Stage 2}: Fix $\bm a =  \hat{\bm a}$, find the optimal SINR requirement $\gamma^*$ using a sequence of linear programs with binary search. Denote the optimal power allocation as $ \hat{\bm p}$. \\
\\
\hline
\end{tabular}
\end{table}


By a simple argument, we readily see that this algorithm indeed solves problem $(\mathrm P_1^{\prime})$ globally.
Suppose $\hat{\gamma}$ is the minimum SINR value corresponding to $(  \hat{\bm p},  \hat{\bm a})$. If $\hat{\gamma} \geq 1$, then the problem $(\mathrm P_1^{\prime})$ is feasible and
$(  \hat{\bm p},  \hat{\bm a})$ is an optimal solution. If $\hat{\gamma} < 1$, then we can prove that problem $(\mathrm P_1^{\prime})$ is infeasible: in fact, $\hat{\bm a}$ is the only candidate for a feasible BS association, but $\hat{\gamma} < 1$ implies that $\hat{\bm a}$ is infeasible, thus
there is no feasible BS association.

Since both Hungarian algorithm and linear programming can be implemented in polynomial time (see \cite{HungarianAlgo} for the analysis of Hungarian algorithm), we have proved that the algorithm presented in Table \ref{Hungarian} is a polynomial time algorithm that solves problem $(\mathrm P_1^{\prime})$ globally.

Next we prove the equivalence of $(\mathrm P_1^{\prime})$ and $(\mathrm P^{\prime})$.
We only need to prove that the QoS constraints ${\rm SINR}_k  \geq 1,\;\forall\;k$ imply that each BS only serves at most one user,
then combining with the fact $K=N$ we obtain that any feasible BS association $\bm a$ must be a permutation of $\{1,\dots, K \}$. 
The following result builds a more general relationship between the maximum number of users one BS can serve and the achievable min-SINR. We refer the readers to Appendix \ref{appen: proof of rate vs num of users} for detailed proof.
\begin{proposition}\label{Prop: rate vs num. of served users}
{\it In a network with $K$ users and $N$ BSs, let $\bm a$ be an arbitrary BS association (not necessarily a one-to-one matching). Suppose the min-SINR achieved by a given tuple $(\bm p, \bm a)$ is no less than $1/m$, i.e.,
$$
{\rm SINR}_k  = \frac{ p_{k} g_{a_k k} }{\sigma_{k}^2 + \sum_{j \neq k}p_{j} g_{a_j k}  } \geq \frac{1}{m}, \quad  k=1,\dots, K,
$$ where $m$ is a positive integer.
 Then $\max_{n}|\Omega_n|\le m$, that is, each BS is associated with at most $m$ users.}   
\end{proposition}
A direct consequence of the above result is that the constraints ${\rm SINR}_k  \geq 1$ imply that each BS only serves at most one user. Therefore,
$(\mathrm P_1^{\prime})$ and $(\mathrm P^{\prime})$ are equivalent.
 Q.E.D. 

\begin{rmk}
One interesting finding is that whenever $(\mathrm P'_1)$ is feasible, BS association and power allocation can be done {\it separately}. This is because the feasible BS association only depends on the channel information, not on the power vector (cf. Lemma \ref{key lemma}). However, this finding is based on the special structure of problem  $(\mathrm P_1^{\prime})$, and it is not clear whether it can be generalized to the network with more antennas or with an unequal number of BSs and users.
\end{rmk}

\begin{rmk}
The algorithm in Table \ref{Hungarian} can be used to solve problem $(\mathrm P_1)$, but the global optimality is not guaranteed. If $\hat{\gamma} \geq 1$, then $(  \hat{\bm p},  \hat{\bm a})$ is clearly the globally optimal solution to problem
$(\mathrm P_1)$. If $\hat{\gamma} < 1$, then $(  \hat{\bm p},  \hat{\bm a})$ can be a suboptimal solution.
This algorithm can also be used to solve problem $(\mathrm P)$ when $K=N$ since any feasible solution of $(\mathrm P_1)$ is also feasible for $(\mathrm P)$.
If $\hat{\gamma}$ is higher than one, $(  \hat{\bm p},  \hat{\bm a})$ is also the global optimal solution to problem
$(\mathrm P)$. 
\end{rmk}

The previously proposed algorithms \emph{DLSum} and \emph{DLSumA} are designed as suboptimal algorithms for problem $(\mathrm P)$.
An interesting finding is that similar to the algorithm in Table \ref{Hungarian}, \emph{DLSum} and \emph{DLSumA} also globally solve a subclass of problem $(\mathrm P)$, i.e.
  $(\mathrm P)$ for the case $K=N$ if the produced min-SINR is higher than one;
   see a precise statement in the following Proposition \ref{prop: DLSum and optimal gamma}.
We relegate the proof of Proposition \ref{prop: DLSum and optimal gamma} to Appendix \ref{appen: proof of DLSum also globally optimal}.

  \begin{proposition}\label{prop: DLSum and optimal gamma}
{\it Consider a network with an equal number of BSs and users, i.e. $K=N$, and equal noise power $\sigma_k^2 = \sigma^2$.
 Suppose the optimal objective value of  $(\mathrm P)$ is $\gamma^*$ and the min-SINR achieved by \emph{DLSum} is $\gamma^{\rm DL}$.
 Then we have:
 \begin{equation}\label{gamma relation 1}
 \gamma^* \geq 1 \Longleftrightarrow \gamma^{\rm DL} \geq 1 ;
 \end{equation}
 \begin{equation}\label{gamma relation 2}
 \gamma^{\rm DL} \geq 1 \Longrightarrow \gamma^* = \gamma^{\rm DL}.
 \end{equation}
 If $\gamma^{\rm DL}$ is defined as the min-SINR achieved by \emph{DLSumA}, both \eqref{gamma relation 1} and \eqref{gamma relation 2} still hold.
  }
\end{proposition}

Proposition \ref{prop: DLSum and optimal gamma} implies that \emph{DLSum} and \emph{DLSumA} both solve $(\mathrm P^{\prime})$
to global optima.
Note that although \emph{DLSum} and \emph{DLSumA} are very efficient in numerical experiments, they are not known to be polynomial time algorithms
  (in fact, they are pseudo-polynomial time algorithms; see \cite{SunUL} for the analysis for a similar algorithm).



\subsection{\emph{AUFP}: A Semi-distributed Algorithm}
So far, we have proposed \emph{DLSum} (and its variant \emph{DLSumA}) and the algorithm in Table \ref{Hungarian}, which can solve $(\mathrm P^{\prime})$ (also $(\mathrm P_1^{\prime})$) globally. However, they require a central controller that knows all the channel information. In this section, we propose a semi-distributed algorithm that solves $(\mathrm P_1^{\prime})$ globally.  

The framework of our semi-distributed algorithm is the same as the algorithm in Table \ref{Hungarian}: solving an assignment problem with gains $\{ \log g_{ij} \}$
in Stage 1, and computing the optimal power allocation with fixed BS association in Stage 2.
The difference is that in Stage 1 we replace the Hungarian algorithm with the auction algorithm \cite{Bert88,Bert92}, while in Stage 2 we use the fixed point algorithm described in \eqref{IBC fixed point}.
Since we require $\bm a$ to be a one-to-one matching, the norm $\| \cdot \|_{\Omega}$ defined in \eqref{Omega norm def} becomes
\begin{equation}\label{Omega norm def, simple}
\begin{split}
\| \bm p \|_{\Omega} = \max_k \frac{ p_k }{ \bar{p}_{a_k} }.
\end{split}
\end{equation}
The proposed two-stage algorithm, referred to as \emph{AUFP} (Auction-Fixed-Point algorithm), is presented in Table \ref{table of Auction}.

\begin{table}[htbp]
\caption{ \emph{AUFP}: Distributed Algorithm for Joint BS Association and Power Allocation}
\label{table of Auction}
\normalsize
\begin{tabular}{p{460pt}}
\hline
\\
\textbf{Stage 1}: Compute a BS association $\bm a$.\\
Initialization: Fix $\epsilon > 0$. Pick random price vector $(w_1,\dots,w_N)$ (not necessarily positive). \\ 
Repeat until convergence: \\
1) Bidding Phase: \\
 $\quad $  For each unassigned user $k $: \\
 $\quad $ 1.1) Find a BS $b_k \triangleq \arg \max_n \{\log(g_{n k})- w_{n} \} $. \\
$\quad  $ 1.2) Compute the bidding increment    \\
  $ \quad \quad \gamma_k = (\log(g_{b_k k})- w_{b_k} ) - \max_{n \neq b_k} \{\log(g_{n k})- w_{n} \} .$ \\

2) Assignment Phase: \\
 $\quad $ For each BS $n $ that $U_n \triangleq \{k: b_k =n \}$ is nonempty: \\
$\quad $ 2.1) Find the highest bidder $j = \arg \max_{k \in U_n} \gamma_k  $.   \\
$\quad  $ 2.2) Update BS association: \\
$\quad \quad \quad \; $ The user assigned to BS $n$ becomes unassigned;  \\
 $\quad \quad \quad \; $ $a_j  \longleftarrow n $. \\
$\quad  $ 2.3) Update price: $w_n \longleftarrow w_n + \gamma_j + \epsilon$. \\
\\
\textbf{Stage 2}: Given $\bm a$, compute a power vector $\textbf{p}$. \\
Initialization: pick random power vector $\textbf{p}(0)$. \\
Loop $t$: \\
$\quad  \quad \quad $ $\bm p(t+1) \longleftarrow  \frac{ M( \bm p(t)) }{ \| M( \bm p(t)) \|_{\Omega} },$ \\
$\; $ where $ M(\cdot) $ is defined in \eqref{def of M}, $\| \cdot \|_{\Omega}$ is defined in \eqref{Omega norm def, simple}. \\
\\
\hline
\end{tabular}
\end{table} 

For the assignment problem with gains $\{\log(g_{ij})\}$, the auction algorithm will terminate in $\frac{\max_{ij} |\log g_{ij}| }{\epsilon}$ iterations if the initial prices are all zeros \cite{Bert92}.
Moreover, the total gain of the final assignment computed by the auction algorithm differs from the optimal value by at most $K\epsilon$ \cite[Proposition 1]{Bert92}. Therefore the auction algorithm converges to the optimal solution of the assignment problem if $\epsilon$ is small enough.

There are several variants of the auction algorithm \cite{Bert92}.
One variant is the Gauss-Seidel version, in which a single unassigned user bids at each iteration; in contrast, the version presented here is the Jacobi version where all unassigned users bid at each iteration.
Another variant, called $\epsilon$-scaling, executes the auction algorithm for several rounds where decreasing values of $\epsilon$ are used at each round.
The auction algorithm can also be applied to the sparse problems, in which each user is only allowed to be matched with a given subset of persons.
The BS association problem is an example of the sparse problem since typically each user is only allowed to be served by several nearby BSs.
Introducing the sparse structure will simplify the auction algorithm, but also make it necessary to detect infeasibility (i.e. the case that no legal one-to-one matching exists).
We refer interested readers to \cite{Bert92} for more details of these variants.

Next we illustrate how the auction algorithm can be implemented in a distributed manner.
Let $U$ denote the set of BSs that receive at least one bid.
At the end of the assignment phase, each BS $n \in U$ updates its ``price'' $w_n$ and broadcasts $e^{-w_n/2}$. Since the price $w_n$ is nondecreasing throughout the algorithm, the transmit power budget constraint $e^{-w_n} \leq \bar{p}_n$ is satisfied if the initial value of $w_n$ is at least $-\log \bar{p}_n$.
 In the bidding phase, each user $k$ receives a signal from each BS $n \in U$ with signal power $g_{nk}e^{-w_n} = e^{\log(g_{n k})- w_{n}} $. Each unassigned user $k$ finds the BS $b_k $  providing the maximal value and the bidding increment $\gamma_k$ based on the new information $g_{nk}e^{-w_n}, n \in U$ and previously stored information $g_{m k}e^{-w_m}, m \notin U$ (cf. Step 1.2).
User $k$ then sends $\gamma_k$ to its intended BS $b_k$.
In the assignment phase, let $U$ now denote the new set of BSs that receives bids.
Suppose each BS $n$ knows the local channel information $g_{nk}, \;\forall\; k$.
Each $n \in U$ receives multiple bidding increments $\gamma_k$'s and finds the largest one $\gamma_j$. BS $n$ then notifies the previously associated user and the new associated user $j$ the change of assignment. As stated earlier, BS $n$ also updates $w_n$ and broadcasts $e^{-w_n/2}$.

As mentioned in Section \ref{subsec: fixed BS, update power}, the fixed-point algorithm for power allocation in Stage 2 converges to the optimal power vector (for the fixed BS association $\hat{\bm a}) $ at a geometric rate.
The fixed-point algorithm in Stage 2 can be implemented in a semi-distributed fashion.
In particular, computing $M_k(\bm p) =  (   \sigma_{k}^2 + \sum_{i \neq k} p_i g_{a_i k} )/ g_{a_k k}  = \mathrm{SINR}_k/p_k$ only requires local measurement
of SINR and the power $p_k$ computed in the last iteration. Computing $\| M(\bm p)\|_{\Omega} = \max_k ( M_k(\bm p) /\bar{p}_{a_k} )$ requires comparison of the information $M_k(\bm p) /\bar{p}_{a_k}$ from each user $k$, which can be executed in a central controller. 
Since no global channel information is required, this algorithm can be viewed as a semi-distributed algorithm.

In summary, \emph{AUFP} can be used to solve problem $(\mathrm P_1^{\prime})$ in a semi-distributed fashion, and
it finds the global optima of $(\mathrm P_1^{\prime})$ if $\epsilon$ is small enough.
\emph{AUFP} can also be viewed as an algorithm to solve $(\mathrm P_1)$ and the original problem $(\mathrm P)$ when $K=N$. One can easily verify that (\ref{gamma relation 1}) and (\ref{gamma relation 2}) still hold if we replace $\gamma^{\rm DL}$ by $\gamma^{\rm AUFP}$ (the min-SINR achieved by \emph{AUFP}). A direct consequence is that if $\gamma^{\rm AUFP} \geq 1,$ then $\gamma^{\rm AUFP}$ is the optimal value of both $(\mathrm P_1)$ and $(\mathrm P)$ when $K=N$.


%
%

\section{Simulation Results}\label{sec:simulation}
In this section, we evaluate the performance of the proposed algorithms. 
Consider a HetNet that consists of $N_{\mathrm m}$ hexagon macro cells, each containing one macro BS in the center. The distance between adjacent macro BSs is 1000m. There are $\beta$ pico BSs randomly placed in each macro cell (at least $250$m apart from the central macro BS), thus in total there are $N = (\beta +1) N_{\mathrm m}$ BSs.
Suppose the maximum power of each macro BS is $16$dB higher than that of each pico BS.
 Suppose the noise power at each user is a constant $ \sigma^2 = 1$ and the maximum power of each pico BS is $P_{\rm pico}$.
 Define the signal to noise ratio as $\text{SNR} = 10 \log_{10}(P_{\rm pico})$.
 The channel gain from BS $n$ to user $k$ at a distance $d_{nk}$ is $g_{nk} =
S_{nk}(200/d_{ik})^{3.7} $, where $10 \log_{10} S_{i,k} \sim \mathcal{N}(0,64)$ models the shadowing effect.
 There are $K $ mobile users in the network, and we consider two kinds of user distributions:
 \begin{itemize}
 \item Congested: $\lfloor \sqrt{K} \rfloor$ users are placed randomly in one macro cell, while other users are uniformly distributed in the network area.
 \item Uni-in-cell: {\color{black}
    Each user $k$ is uniformly randomly placed in cell $\phi(k)$, where $\phi$ is a periodic function with period $N$, i.e. $\phi_{k} = \phi_{k - N}, \forall k$ and $( \phi(1), \dots, \phi(N) )$ is a random permutation of $(1,2,\dots, N)$. 
    For instance, if $K = N$, then one user is randomly placed in each of the $N$ cells (including both macro cells and pico cells);
    if $K = 2N$, then two users are randomly placed in each of the $N$ cells.
 This setting models the practical scenario that the number of users that are scheduled to be served in each cell in one time/frequency slot is
 usually close to a constant.
      }
 \end{itemize}
Each point of the plots in this section is obtained by averaging over $500$ Monte Carlo runs.

\begin{figure}[htbp]
    \vspace{-.1cm}
    \centering
{\includegraphics[width=0.5\linewidth]{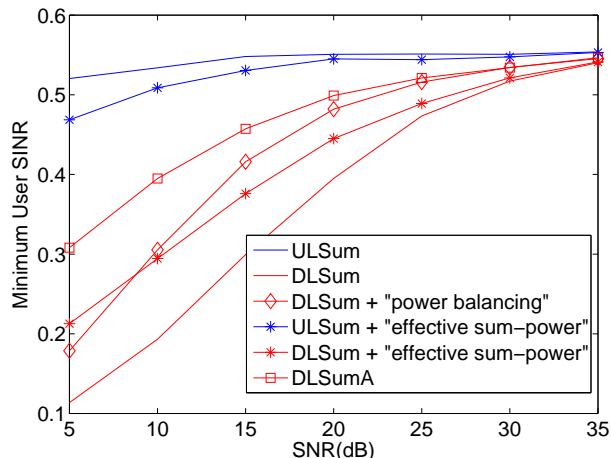} }
\vspace{-.5cm}
  \caption{
  \footnotesize{  Comparison of the minimum user SINR achieved by \emph{ULSum}, \emph{DLSum} and their variants, including \emph{ULSum} with ``power balancing' (\emph{ULSumA}), \emph{DLSum} with both ``effective sum-power'' and ``power balancing'' (\emph{DLSumA}). $N_{\mathrm m}= 16$ macro cells, $\beta = 2$ pico BSs in each macro cell, $N = 48$ BSs and $K = 75$ users. }
}\label{Fig1sum} \vspace{-.1cm}
\end{figure}

In the first experiment, we examine the effectiveness of the two techniques  ``power balancing'' and ``effective sum-power'' proposed in section \ref{two tech for Algo 2}. 
Specifically, we consider six algorithms: \emph{ULSum}, \emph{DLSum},
\emph{DLSum} combined with ``effective sum-power'', and these algorithms
combined with ``power balancing''. Note that \emph{ULSum} with ``power balancing'' is exactly \emph{ULSumA}, and \emph{DLSum} with both ``effective sum-power'' and ``power balancing'' is \emph{DLSumA}.
We test these algorithms for the user distribution ``Uni-in-cell'', and consider $N_{\mathrm m}= 16$ macro cells, $\beta = 2$ pico BSs in each macro cell, $N = 48$ BSs and $K = 75$ users.
Fig. \ref{Fig1sum} shows that \emph{ULSum} with ``power balancing'' provides a tighter upper bound than \emph{ULSum}, especially in low SNR regime.
Among the four variants of \emph{DLSum}, the one combining both techniques (\emph{DLSumA}) provides the highest minimum SINR, thus the proposed two techniques
improve the performance of \emph{DLSum}.


\begin{figure}[htbp]
    \vspace{-.1cm}
    \centering
{\includegraphics[width=0.5\linewidth]{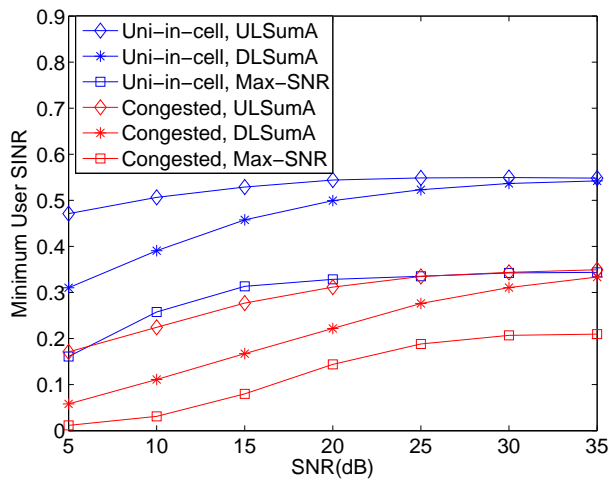} }
\vspace{-.5cm}
  \caption{
  \footnotesize{  Comparison of the minimum user SINR achieved by \emph{ULSumA}, \emph{DLSumA} and \emph{Max-SNR}, for two user distributions ``Uni-in-cell''
  and ``Congested''. $N_{\mathrm m}= 16$ macro cells, $\beta = 2$ pico BSs in each macro cell, $N = 48$ BSs and $K = 75$ users. }
}\label{Fig2_maxSNR} \vspace{-.3cm}
\end{figure}

In the second experiment, we compare the performance of the proposed algorithm \emph{DLSumA} with the upper bound computed by \emph{ULSumA} (i.e. the optimal value of the sum power constrained problem after channel scaling) under different scenarios.
 We also compare them with the algorithm \emph{Max-SNR}, which computes the BS association based on the maximum receive SNR, i.e. $a_k = \arg \max_n \{g_{nk}\bar{p}_{n} \} $.
For a fair comparison, the optimal power allocation corresponding to \emph{Max-SNR} is then computed by the fixed point iteration \eqref{IBC fixed point}.
We test the three algorithms for ``Uni-in-cell'' and ``Congested''; other settings are the same as in the first experiment.
Fig. \ref{Fig2_maxSNR} shows that the gap between \emph{DLSumA} and the upper bound is very small when ${\rm SNR} \geq 25$dB, i.e. \emph{DLSumA} is nearly optimal in high SNR regime. The performance gap between \emph{DLSumA} and \emph{ULSumA} is larger for the user distribution ``Congested'' than for ``Uni-in-cell'', which is reasonable since it is difficult to deal with congested networks.
Note that the optimal curve should lie between the curves of \emph{DLSumA} and \emph{ULSumA}, thus we can tell approximately how far \emph{Max-SNR} is from the true optima: in all cases, the optimal value is at least $50 \%$ higher than the value achieved by \emph{Max-SNR}; in most cases, the performance gain is no more than $100 \%$.


\begin{figure}[htbp]
    \vspace{-.1cm}
    \centering
{\includegraphics[width=0.5\linewidth]{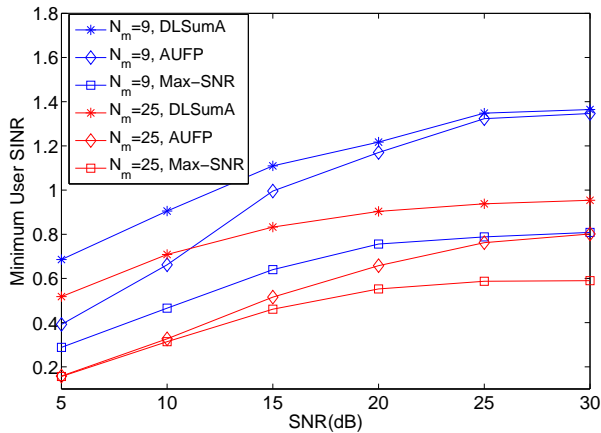} }
\vspace{-.5cm}
  \caption{
  \footnotesize{  Comparison of the minimum user SINR achieved by \emph{DLSumA}, \emph{AUFP} and \emph{Max-SNR}. $N_{\mathrm m}= \{9, 25\}$ macro cells,
   $\beta = 1$ pico BS in each macro cell, $N = \{18, 50 \}$ BSs and $K = N$ users. }
}\label{Fig3auc} \vspace{-.1cm}
\end{figure}

In the third experiment, we evaluate the performance of the semi-distributed algorithm \emph{AUFP} by comparing it with the centralized algorithm \emph{DLSumA} and the distributed algorithm \emph{Max-SNR}.
We test these algorithms for the user distribution ``Uni-in-cell'' with the following parameters: $N_{\mathrm{m}} = \{9, 25\}$ macro cells, $\beta = 1$ pico BS per macro cell, $N = \{ 18, 50 \}$ BSs and $K = \{ 18, 50 \}$ users.
We show two figures for this experiment. Figure \ref{Fig3auc} compares the average performance of these methods at different SNR levels.
 It shows that for both $N_{\mathrm m}= 9$ and $N_{\mathrm m}= 25$, \emph{AUFP} significantly outperforms \emph{Max-SNR} in high SNR regimes, and has almost the same performance as \emph{DLSumA} when SNR is higher than $25$ dB; however, in low SNR regimes \emph{AUFP} performs worse than \emph{DLSumA}.
These facts indicate that the performance of \emph{AUFP} highly depends on the SNR level of the system.

\begin{figure}[htbp]
    \vspace{-.1cm}
    \centering
{\includegraphics[width=0.5\linewidth]{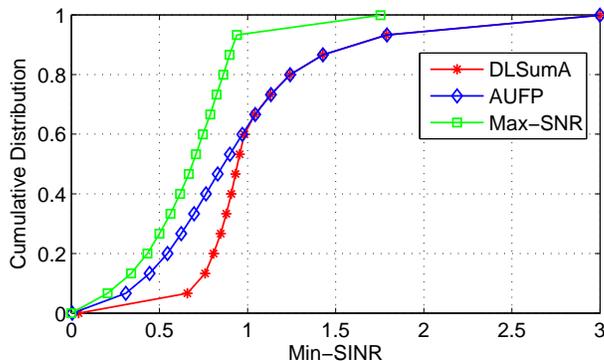} }
\vspace{-.5cm}
  \caption{
  \footnotesize{ CDFs (Cumulative Distribution Function) of the min-SINR achieved by \emph{DLSumA}, \emph{AUFP} and \emph{Max-SNR}. SNR = $15$dB, $N_{\mathrm m}= 9$ macro cells,   $\beta = 1$ pico BS in each macro cell, $N = 18$ BSs and $K = 18$ users.   }
}\label{Fig3CDF} \vspace{-.1cm}
\end{figure}

Besides the average performance, we are also interested in the distribution of the obtained min-SINR.
In Figure \ref{Fig3CDF} we plot the CDF (Cumulative Distribution Function) of the min-SINR of these methods for the case $N_m = 9, {\rm SNR} = 15 \rm dB$.
This plot is obtained through 5000 Monte Carlo runs.
For less than 5$\%$ of the runs, the achieved min-SINR is higher than 3;  we set the min-SINR values of these cases to 3 in order to make the figure more readable.
Figure \ref{Fig3CDF} shows that \emph{AUFP} and \emph{DLSumA} achieve exactly the same performance when the min-SINR is higher than $1$. This phenomenon matches our theoretical results that both \emph{AUFP} and \emph{DLSumA} achieve $\gamma^*$, the global optimal value of $(\rm P)$, as long as $\gamma^* \geq 1$.
We also find that \emph{max-SNR} rarely achieves a min-SINR value that is higher than 1. This phenomenon may be explained as follows: \emph{max-SNR} usually associates $2$ or more users to one BS, in which case the achieved min-SINR must be smaller than 1 according to Proposition \ref{Prop: rate vs num. of served users}.


\section{Concluding Remarks}
\label{Conclusion}
In this paper, we systematically investigate the max-min fairness problem by joint BS association and power allocation in a downlink cellular network. We consider two specific types of BS association strategies, analyze their computational complexity, and design efficient algorithms. In particular, we show that for the general case where one BS can serve multiple users, the problem is NP hard. We propose a fixed point algorithm to compute an upper bound of the optimal value, as well as a two-stage fixed point algorithm to solve the original problem, which computes a lower bound. The proposed bounds can be used to evaluate other algorithms, such as the traditional approach that determines BS association based on the maximum receive SINR. We further show that the problem of finding a one-to-one matching between users and BSs is still NP-hard in general, but becomes polynomial time solvable after adding certain SINR constraints. We propose to use the auction algorithm to compute the user-BS matching in this case. One future direction is to solve the problem with multiple time/frequency slots by joint BS association, power control and scheduling.  Another interesting direction is to combine BS association with SDNs (Software Defined Networking) \cite{SDNsurvey,Liao13SDN}.


\text{ }\newline
\appendix
\par\noindent

\subsection{NP Hardness Proof of Theorem \ref{theoremNP} }\label{appen: proof of NP hard}
Theorem \ref{theoremNP} is proved based on a polynomial time transformation from
the 3-SAT problem, which is a known NP-complete problem
\cite{garey79}. The 3-SAT problem is described as follows. Given a conjuctive formula $S = C_1 \wedge \dots \wedge C_M$ defined on $T$ Boolean variables
$X_1\cdots,X_T$, where $C_m= Y_1\vee Y_2\vee Y_3$ with
$Y_i\in\{X_1,\cdots,X_T, \bar{X}_1,\cdots,\bar{X}_T\}$, the problem
is to check whether $S$ is satisfiable, i.e. whether
 there exists a truth assignment for the Boolean
variables such that all clauses $C_m$ are satisfied.

\begin{figure*}[htb] \vspace*{-.3cm}
    \begin{minipage}[t]{0.48\linewidth}
    \centering
    {\includegraphics[width=
    0.9 \linewidth]{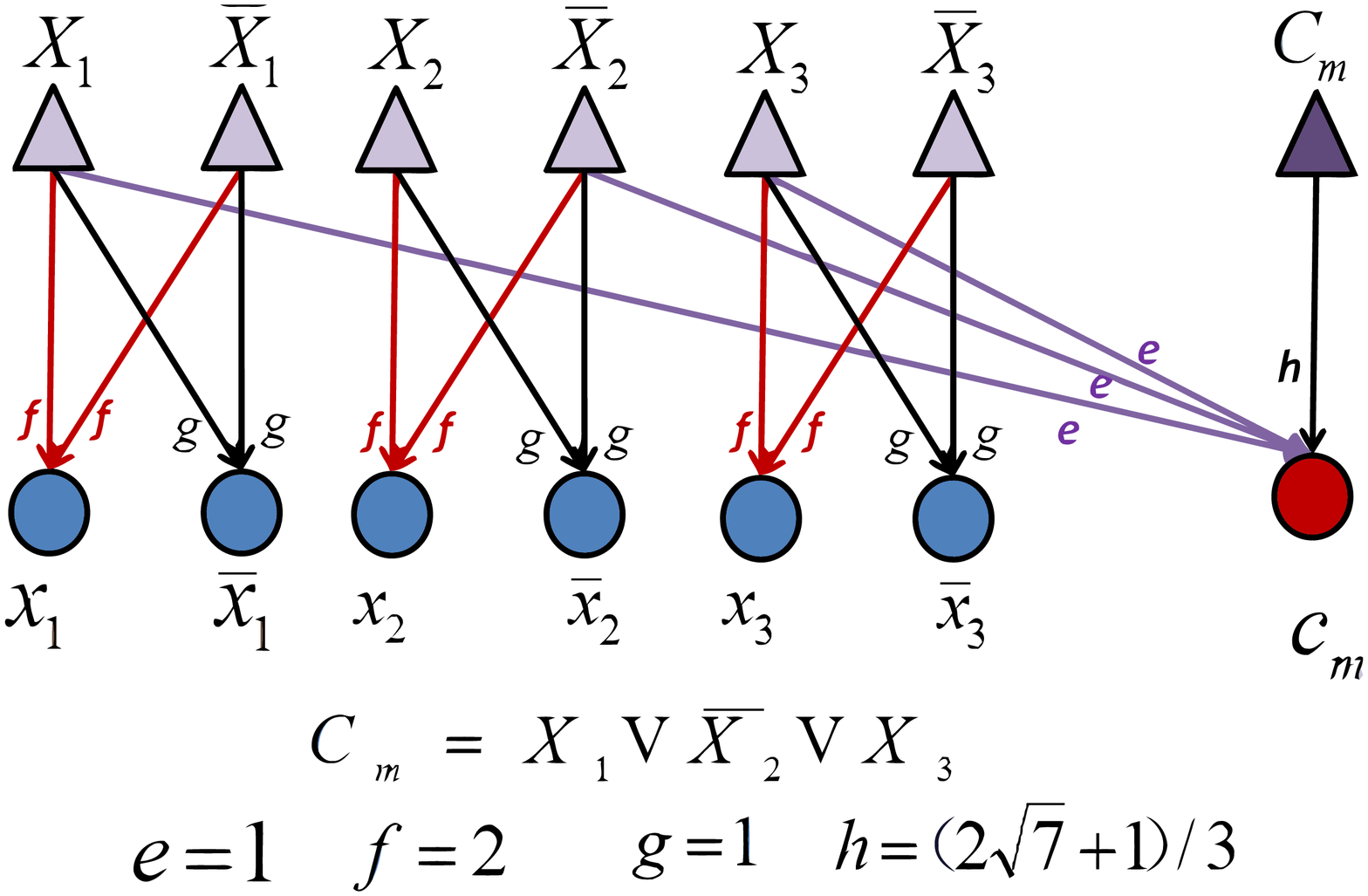}
    \vspace*{-0.2cm}\caption{Construction of the network for clause
    $C_m=X_1 \vee\bar{X}_2 \vee X_3$. }\label{figConstruction}
    \vspace*{-0.1cm}}
\end{minipage}\hfill
    \begin{minipage}[t]{0.48\linewidth}
    \centering
    {\includegraphics[width=
0.9  \linewidth]{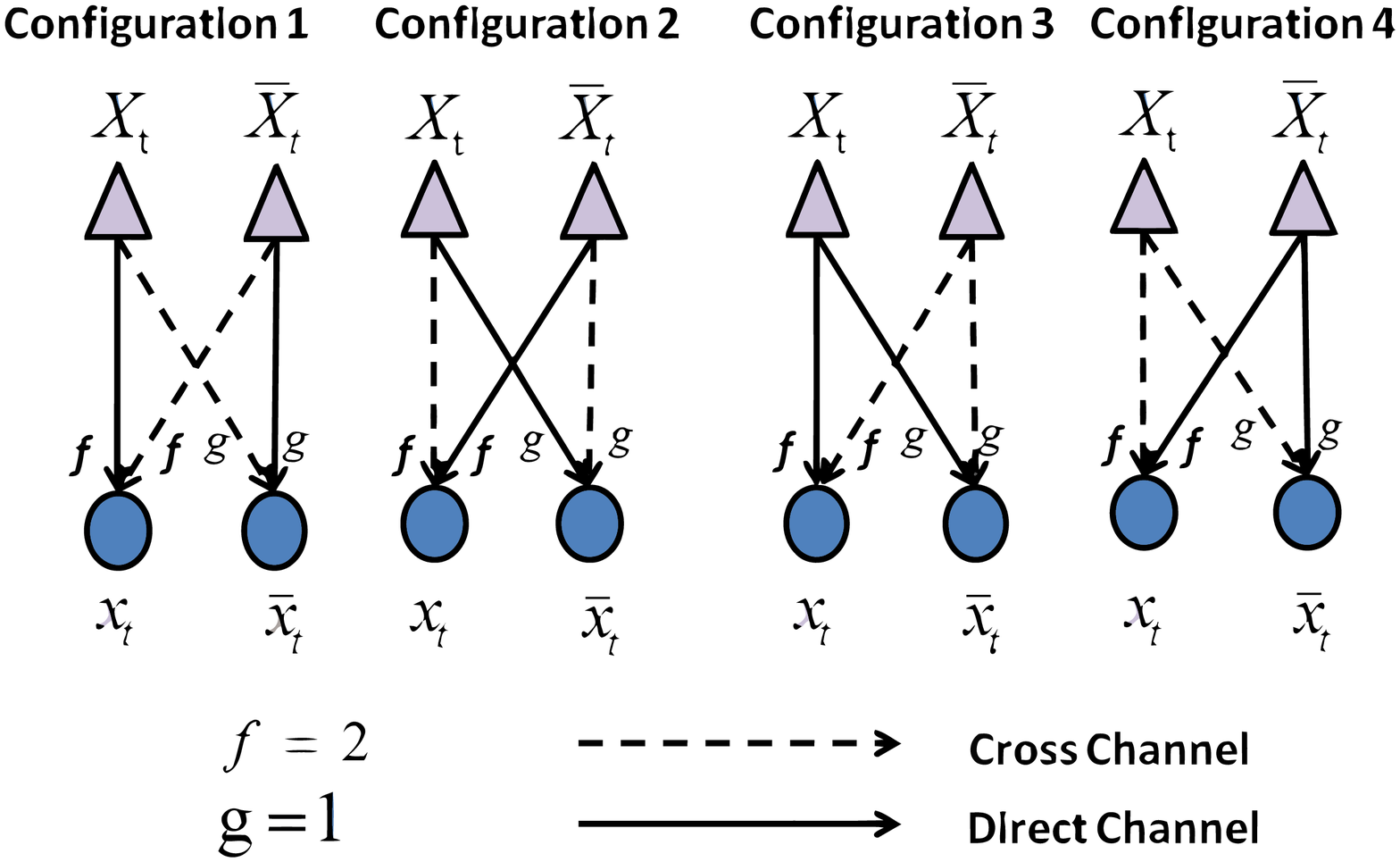}
\vspace*{-0.2cm}\caption{Four possible user-BS association among BSs
$X_t$, $\bar{X}_t$ and users $x_t$, $\bar{x}_t$.
}\label{figConfiguration} \vspace*{-0.1cm}}
\end{minipage}
\vspace*{-0.4cm}
    \end{figure*}

Given any formula $S$ with $M$ disjunctive
clauses and $T$ variables, we construct an instance of multiple BS
multi-user network with $M+2T$ BSs and $M+2T$ users.
Let $\pi(C_m)$ denote the set of terms consisting clause $C_m$, i.e., if
$C_m=\bar{X}_1\vee \bar{X}_2\vee X_4$, then $\pi(C_m)=\{\bar{X}_1,
\bar{X}_2, {X}_4\}$.
In the constructed network, we let $\sigma^2_k=1$, $\bar{p}_n=1$, for all $k,n$.
To illustrate the idea of construction, for one clause $C_m=X_1\vee \bar{X}_2\vee X_3$,
the constructed subnetwork is shown in Fig. \ref{figConstruction}. 
In general,
we construct one {\it clause user} $c_m$ and one {\it clause BS}
$C_m$ for each clause $C_m$; we construct $2$ {\it variable users}
$\bar{x}_t$, ${x}_t$ and $2$ {\it variable BSs} $\bar{X}_t, X_t$ for each variable $X_t$.
The channel gains are set as follows:
\begin{align}
h_{C_m,q}&= \left\{ \begin{array}{ll}
\frac{2\sqrt{7} + 1}{3} ,&~\textrm{if}~q=c_m, \\  
0,&\textrm{otherwise}.
\end{array}\right.
\end{align}
\begin{align}
h_{X_t,q}&= \left\{ \begin{array}{ll}
2,&\textrm{if}~q=x_t,\\
1,&\textrm{if}~q=\bar{x}_t,\\  
1,&\textrm{if}~q=c_m,~\textrm{and}~X_t\in\pi(C_m),\\
0,&\textrm{otherwise}.
\end{array}\right.
\end{align}
\begin{align}
h_{\bar{X}_t, q}&= \left\{ \begin{array}{ll}
1,&\textrm{if}~q={x}_t,\\  
2,&\textrm{if}~q=\bar{x}_t,\\
1,&\textrm{if}~q=c_m,~\textrm{and}~\bar{X}_t\in\pi(C_m),\\
0,&\textrm{otherwise}.
\end{array}\right.
\end{align}
Note that the clause BS $C_m$ has nonzero channels only to the clause users $c_m$; the variable BS
$X_t$ (or $\bar{X}_t$) has nonzero channels only to the variable users $x_t$, $\bar{x}_t$, and the clause users corresponding to the clauses containing variable $X_t$ or $\bar{X}_t$.
As a result, a clause user $c_m$ can only be associated with its
corresponding clause BS $C_m$ or the variable BS
$Y$ that satisfies $Y \in\pi(C_m)$ (e.g. in Fig. \ref{figConstruction}, $c_m$ can only be associated with clause BS $C_m$ or variable BS $X_1, \bar{X}_2, X_3$). A variable user $x_t$ (or
$\bar{x}_t$) can only be associated with BS $X_t$ or $\bar{X}_t$.

For each $t \in \{1,\dots, T \}$, let $\Gamma_t$ denote the subnetwork that consists of BSs $X_t$, $\bar{X}_t$ and users
$x_t$, $\bar{x}_t$ (see Fig. \ref{figConfiguration}).
Let $\{ (Y_{t1}, x_t), (Y_{t2}, \bar{x}_t) \} $ denote the user-BS association that user $x_t$ is
associated with BS $Y_{t1}$ and user $\bar{x}_t$ is associated with BS $Y_{t2}$, where $Y_{t1}, Y_{t2} \in \{X_t, \bar{X_t} \}$.
 Clearly there are four possible user-BS associations for this subnetwork:
\begin{equation}\nonumber
\begin{split}
\text{Configuration 1}: \{ (X_t, x_t), (\bar{X}_t, \bar{x}_t) \}; \\
\text{Configuration 2}: \{ (X_t, \bar{x}_t ), (\bar{X}_t,x_t ) \} ; \\
\text{Configuration 3} (X_t \text{serves both}): \{ (X_t, x_t ), (X_t, \bar{x}_t ) \} ; \\
\text{Configuration 4} (\bar{X}_t \text{serves both}): \{ (\bar{X}_t, x_t ), (\bar{X}_t, \bar{x}_t ) \} ; \\
\end{split}
\end{equation}

The following Lemma \ref{SINR compute lemma} characterizes the max-min solution of $\Gamma_t$. This result implies that an {\it
upper bound} on the min-SINR of the entire
network constructed above is $\gamma^* = (\sqrt{7}-1)/3 $.
The proof of Lemma \ref{SINR compute lemma} is given at the end of this Appendix. 

\begin{lemma}\label{SINR compute lemma}
{\it The optimal min-SINR of the subnetwork $\Gamma_t$ is $\gamma^* = (\sqrt{7}-1)/3 \approx 0.5486 $.
This min-SINR is achieved either in configuration 1 with power $p_{X_t}(1)=(\sqrt{7}-1)/2 \approx 0.8229, p_{\bar{X}_t}(1)=1$ , or in configuration 2 with $p_{X_t}(2)=1,
p_{\bar{X}_t}(2)=(\sqrt{7}-1)/2 \approx 0.8229$.}
\end{lemma}

In the following, we will show that an instance of 3-SAT is satisfiable if and only if
the corresponding network we constructed achieves the min-SINR $\gamma^* = (\sqrt{7}-1)/3 \approx 0.5486 $.  

We first prove the ``only if'' direction. Suppose a formula $S$ is satisfiable, we
need to prove that the min-SINR of the corresponding network is $\gamma^*$.
Suppose $(X_1, \dots, X_T) = (s_1, \dots, s_T)$ is a truth assignment that satisfies all
clauses.
If $s_t = 1$, let the subnetwork $\Gamma_t$ choose configuration $1$,
and use power $p_{X_t} = (\sqrt{7}-1)/2, p_{\bar{X}_t} = 1$ to transmit;
otherwise, let $\Gamma_t$ choose configuration $2$, and use power $p_{X_t} = 1 , p_{\bar{X}_t} = (\sqrt{7}-1)/2 $ to transmit.
Note that by our construction, $p_{Y} = (\sqrt{7}-1)/2$ if $Y \in \{ X_t, \bar{X}_t\}$ evaluates to $1$.
According to Lemma \ref{SINR compute lemma}, the min-SINR of $\Gamma_t$ is $\gamma^* = (\sqrt{7}-1)/3 $.  

For each $m\in \{ 1,\dots, M \}$, we associate user $c_m$ to BS $C_m$, and let BS $C_m$ transmit with maximum power $1$.
We prove that the SINR of user $C_m$ is at least $\gamma^* = (\sqrt{7}-1)/3 $.
Suppose $C_m= Y_1\vee Y_2\vee Y_3$ with $Y_i\in\{X_1,\cdots,X_T, \bar{X}_1,\cdots,\bar{X}_T\}$.
Since the clause $C_m$ is satisfied, there must exist a term $Y \in \pi(C_m)$ that evaluates to $1$; without loss of generality, assume $Y_1 $ evaluates to $1$.
 As mentioned earlier, $p_{Y_1} = (\sqrt{7}-1)/2 $, and $p_{Y_2}, p_{Y_3} \leq 1$.
 By our construction, only the three BSs $Y_1,Y_2, Y_3$ have interference to the clause user $c_m$,
 thus the aggregated interference at user $c_m$ is upper bounded as follows: $$ \sum_{i=1}^3 h_{Y_i c_m} p_{Y_i} = \sum_{i=1}^3  p_{Y_i} \leq (\sqrt{7}+3)/2 .$$
Consequently, the SINR of user $c_m$ is lower bounded as:
\begin{align}\nonumber
\gamma_{c_m}=\frac{1\times h_{C_m, c_m}}{1+  \sum_{i=1}^3 h_{Y_i c_m} p_{Y_i} } \geq \frac{ 2\sqrt{7}+1 }{3( 1+  \frac{\sqrt{7}+3}{2} )} = \frac{\sqrt{7}  - 1}{3}.
\end{align}
Therefore,  each user in the 
network achieves an SINR that is at least $\gamma^* = (\sqrt{7}-1)/3$. Since $\gamma^*$ is
also an upper bound of the optimal min-SINR, we conclude
that $\gamma^*$ is exactly the optimal min-SINR of the network.

We then show the reverse direction. Suppose that the network we
constructed achieves a min-SINR $\gamma^* = (\sqrt{7}-1)/3$, we prove that the
formula $S$ is satisfiable.
To achieve a min-SINR $\gamma^*$, each subnetwork $\Gamma_t$ must choose either configuration 1 or configuration 2 to transmit,
and $\{ p_{X_t}, p_{\bar{X}_t }  \} = \{1, ( \sqrt{7} -1) /2 \}$.
Note that the variable BSs $X_t, \bar{X}_t$ cannot serve any clause user $c_m$, otherwise the min-SINR of $\Gamma_t$ will be strictly les
than $\gamma^*$; thus each clause user $c_m$ must be associated with the clause BS $C_m$.
Define a truth assignment $(X_1, \dots, X_T) = (s_1, \dots, s_T)$ as follows:
\begin{equation}\label{truth_assign}
s_t =
\begin{cases}
1, & \text{ if }  p_{X_t } = \frac{\sqrt{7} -1 }{2}, \\
0, & \text{ else }.
\end{cases}
\end{equation}
We claim that each clause $C_m = Y_1 \vee Y_2 \vee Y_3$ is satisfied under this truth assignment.
In fact, since the SINR of clause user $c_m$ is at least $\gamma^*$, the total interference generated by the variable
users $Y_1, Y_2, Y_3$ is at most $\frac{1\times h_{C_m, c_m}}{\gamma^*}-1=2+ (\sqrt{7}-1)/2 $.
Thus at least one variable BS $Y_i$ transmits with power strictly less than one.
According to \eqref{truth_assign}, $Y_i$ must evaluate to $1$, thus $C_m$ is satisfied.
Q.E.D.

\emph{Proof of Lemma \ref{SINR compute lemma}}:
The max-min fairness solution for the four configurations can be computed explicitly using results in \cite{Blon05}, \cite{MD10} or \cite{TanChiang2009}.
Let $f = 2, g=1$.
For configuration $1$, the max-min fairness problem is
\begin{equation}\label{max min of config 1}
\begin{split}
 \max_{p_{X_t},p_{\bar{X}_t} } &  \min \left\{ \frac{ p_{X_t} f }{1 + p_{\bar{X}_t}  f },  \frac{ p_{\bar{X}_t} g }{1 + p_{X_t}  g } \right\}, \\
\st \;      & \quad  0 \leq p_{X_t} \leq 1, 0\leq p_{\bar{X}_t} \leq 1.  \\
\end{split}
\end{equation}
Define $F \triangleq \begin{bmatrix} 0 & f/f \\ g/g & 0 \end{bmatrix} = \begin{bmatrix} 0 & 1 \\ 1 & 0 \end{bmatrix} $ that consists of cross-link channel gains scaled by direct-link channel gains, and define scaled noise power vector $v \triangleq (1/f, 1/g)$.
Let $\rho(A)$ denote the largest eigenvalue of $A$, and define $e_1 = (1,0), e_2 = (0,1)$.
By \cite[Theorem 9]{MD10} or \cite[Theorem 2]{TanChiang2009}, the optimal min-SINR of configuration 1 is
given by
\begin{equation}
\begin{split}
& \gamma(1) = \frac{1}{\rho ( F + v e_i^T ) }, \\
\text{where } \quad & i = \arg \min_{j \in \{ 1, 2 \} }  \frac{1}{\rho ( F(1) + v e_j^T ) },
\end{split}
 \end{equation}
and the optimal power vector $ (p_{X_t}(1),p_{\bar{X}_t}(1) ) $ is the eigenvector corresponding to the largest eigenvalue of $ F + v e_i^T$.
It is easy to verify that since $f > g $, we have $i = 2$,  $\gamma(1) = \frac{2}{1/g + \sqrt{1/g^2+ 4(1+1/f)} } = \frac{\sqrt{7} -1 }{3}$, and $ p_{X_t}(1)  = 1/\gamma(1) - 1/g  = (\sqrt{7}-1)/2
 \approx  0.8229 , p_{\bar{X}_t}(1) = 1 $.

Similarly, the optimal min-SINR of configuration 2 is $\gamma(2) = (\sqrt{7}-1)/3$, and the optimal power $ p_{X_t}(2)  = 1, p_{\bar{X}_t}(2)  = (\sqrt{7}-1)/2
 \approx  0.8229 $.

 For configuration 3, the max-min fairness problem is
\begin{equation}\label{max min of config 3}
\begin{split}
 \max_{\bm p_{x_t},p_{\bar{x}_t} } &  \min \left\{ \frac{ p_{x_t} f }{1 + p_{\bar{x}_t}  f },  \frac{ p_{\bar{x}_t} g }{1 + p_{x_t}  g } \right\}, \\
\st \;      & \quad  p_{x_t} + p_{\bar{x}_t} \leq 1,  \\
\end{split}
\end{equation}
 where $p_{x_t}$ and $p_{\bar{x}_t}$ are the power of BS $X_t$ used to serve user $x_t$ and $\bar{x}_t$ respectively.
Problem \eqref{max min of config 3} has the same channel gain matrix $F = \begin{bmatrix} 0 & 1 \\ 1 & 0 \end{bmatrix} $ as \eqref{max min of config 1},
but the difference is that \eqref{max min of config 3} has a sum power constraint.
According to \cite[Theorem 7]{MD10}, the optimal min-SINR of configuration 3 is
$$ \gamma(3) = \frac{1}{ \rho ( F(1) + v e_1^T + v e_2^T ) } =\frac{1}{1/f + 1/g + 1} = 0.4 .$$
Similarly, the optimal min-SINR of configuration 4 is $\gamma(4)  = 0.4$.

Combining the results above, the optimal min-SINR of the subnetwork $\Gamma_t$ is $$ \gamma^* = \max_{1\leq i \leq 4} \gamma(i) = \gamma(1) = \frac{\sqrt{7} -1 }{3} , $$
and $\gamma^*$ is achieved either in configuration 1 with power $p_{X_t}(1)=(\sqrt{7}-1)/2 , p_{\bar{X}_t}(1)=1$ , or in configuration 2 with $p_{X_t}(2)=1,
p_{\bar{X}_t}(2)=(\sqrt{7}-1)/2 $.

\subsection{ Proof of Proposition \ref{Prop: rate vs num. of served users} }\label{appen: proof of rate vs num of users}
Define $\Omega_n = \{k \mid a_k = n \}$, then we need to prove $\max_n |\Omega_n| \leq m$.
Assume the contrary, that $\max_n |\Omega_n| > m+1$.
Without loss of generality, suppose $|\Omega_1| = \max_n |\Omega_n|$, and $ \{1,\dots, m+1 \} \subseteq \Omega_1 $.
Since the min-SINR is no less than $1/m$, we have
\begin{equation}\nonumber
\begin{split}
  & \mathrm{SINR}_k = \frac{ p_{k} g_{a_k k} }{\sigma_{k}^2 + \sum_{j \neq k}p_{j} g_{a_j k}  } \geq \frac{1}{m}, \quad \forall\ k, \\
  \Longrightarrow \; &  p_{k} g_{a_k k} \geq \frac{ \sigma_{k}^2 + \sum_{j \neq k}p_{j} g_{a_j k}}{m} ,\quad \forall\ k.
\end{split}
\end{equation}
For $k\in \Omega_1$, we have $a_k = 1$, thus
\begin{equation}\label{each p is large enough}
\begin{split}
  &  p_{k} g_{1 k} > \frac{  \sum_{j \neq k, 1\leq j \leq m+1} p_{j} g_{1 k} }{m}, \quad k=1,\dots, m+1.  \\
\Longrightarrow \; &  p_{k}  > \frac{  \sum_{j \neq k, 1\leq j \leq m+1} p_{j}  }{m}, \quad k=1,\dots, m+1.
\end{split}
\end{equation}
Without loss of generality, assume $ p_1  = \min_{1\leq k \leq m+1} p_k  $. This implies
$p_1 \leq (\sum_{j =2}^{m+1} p_{j}) /m$, which contradicts \eqref{each p is large enough}. Q.E.D.

\subsection{ Proof of Lemma \ref{key lemma} }\label{appen: proof of assignment lemma}
Without loss of generality, assume $a_k^* = k, k=1,\dots,K.$
Since $\bm{a}^*$ is feasible, there exists a power vector $\bm{p}$ such that
$ \frac{ p_{k} g_{a_k^* k} }{\sigma_{k}^2 + \sum_{j \neq k}p_{j} g_{a_j^* k}  } \geq 1, \forall k. $
Therefore, we have
\begin{equation}\label{1st inequ}
p_{k} g_{k k}  > \sum_{j\neq k}p_j g_{j k }, k=1,\dots, K.
\end{equation}

Consider any assignment $\bm{a}= (a_1,\dots, a_K) \neq \bm{a}^*.$
If $a_k \neq k$, from (\ref{1st inequ}) it follows that $p_{k} g_{k k}  >  p_{a_k} g_{a_k k}$; 
if $a_k = k,$ we have $p_{k} g_{k k}  = p_{a_k} g_{a_k k}$. Combining the two cases, we obtain
\begin{equation}\label{2nd inequ}
p_{k} g_{k k}  \geq p_{a_k} g_{a_k k}, \forall\ k.
\end{equation}
The equality holds if and only if $a_k = k$.
Since there exists some $j$ s.t. $a_j \neq j$, inequality (\ref{2nd inequ}) is strict for $k=j$.

Multiplying (\ref{2nd inequ}) for $k=1,\dots, K$, we have
\begin{equation}\label{3rd ineq}
\Pi_{k=1}^K (p_{k} g_{k k}) > \Pi_{k=1}^K ( p_{a_k} g_{a_k k}).
\end{equation}

Note that $\Pi_{k=1}^K p_{k} = \Pi_{k=1}^K p_{a_k} >0$, thus
\begin{equation}\label{main inequality0}
\Pi_{k=1}^K g_{k k} > \Pi_{k=1}^K g_{a_k k}.
\end{equation}

Taking logarithm of both sides, we get
\begin{equation}\label{main inequality}
\begin{split}
\sum_{k=1}^K \log( g_{kk}) & > \sum_{k=1}^K \log( g_{a_k k }), \\
\text{ for any permutation } &  \bm{a}=(a_1,\dots,a_K) \neq \bm{a}^*.
 \end{split}
\end{equation}

The inequality (\ref{main inequality}) implies that $\bm{a}^* = (1,\dots, K)$ is the unique optimal solution to the assignment problem with gains $\{\log(g_{ij})\}$, thus Lemma \ref{key lemma} is proved. Q.E.D.

\subsection{ Proof of Proposition \ref{prop: DLSum and optimal gamma} }\label{appen: proof of DLSum also globally optimal}
We first prove the following claim.
\begin{claim}\label{claim: DLSum globally solve the matching problem}
{\it Consider a network with an equal number of BSs and users, i.e. $K=N$, and equal noise power $\sigma_k^2 = \sigma^2$.
 Suppose the optimal objective value of  $(\mathrm P)$ is $\gamma^*$. If $\gamma^* \geq 1$, then both algorithms DLSum and DLSumA achieve $\gamma^*$. }
\end{claim}

\emph{Proof:}
Suppose the optimal value of the sum power constrained problem $\rm P_{sum}$ is $\gamma_{\rm sum}$, then $\gamma_{\rm sum} \geq \gamma^* \geq 1$.
By the same argument of Lemma \ref{key lemma} we can show that $\bm a^{\rm UL}$ (corresponding to $\gamma_{\rm sum}$) must be the unique solution to the assignment problem with gains $\{\log g_{ij}\}$.
According to Lemma \ref{key lemma}, $\bm a^{\rm UL}$ must be the optimal BS association for problem $(\mathrm P)$.
Since Stage 2 of \emph{DLSum} solves problem $(\mathrm P)$ with fixed association $\bm a^{\rm UL}$ to global optima, \emph{DLSum} achieves $\gamma^*$.

Next we consider the algorithm \emph{DLSumA}. The scaling of channel gains in Step 0 does not affect the optimal value of $(\mathrm P)$, thus without
loss of generality we just assume $\bar{p}_n = \bar{p}_{\rm max}$.
Under this assumption, in Step 0 we just scale each channel gain by $1$, then Step 1 and Step 2 of \emph{DLSumA} are the same as \emph{DLSum}.
We have proved that $(\bm p^{\mathrm{DL}}, \bm a^{\rm UL})$ computed in Step 1-2 (i.e. by \emph{DLSum}) is an optimal solution to problem $(\mathrm P)$.
In Step 3, the problem with the new sum power constraint $\mathrm P^{\mathrm{UL}}_{\mathrm{sum}} ( g_{nk}, p^{\mathrm{DL}}_n )$
still has an optimal value that is no less than 1 since
\begin{equation}\nonumber
\begin{split}
val( \mathrm P^{\mathrm{UL}}_{\mathrm{sum}} ( g_{nk}, p^{\mathrm{DL}}_n ))  & = val( \mathrm P_{\mathrm{sum}} ( g_{nk}, p^{\mathrm{DL}}_n )) \quad \;
\text( Proposition \ref{prop 2}) \\
& \geq val( \mathrm P(g_{nk}, p^{\mathrm{DL}}_n ) )  \quad \quad \; ( P^{\mathrm{UL}}_{\mathrm{sum}}
( g_{nk}, p^{\mathrm{DL}}_n ) \text{ is a relaxed version of  } P(g_{nk}, p^{\mathrm{DL}}_n )  )  \\
& = val( \mathrm P(g_{nk}, \bar{p}_{\rm max} ) ) \quad \quad  ((\bm p^{\mathrm{DL}}, \bm a^{\rm UL}) \text{ computed by \emph{DLSum} is optimal to }  (\mathrm P) )  \\
& = \gamma^* \geq 1.
\end{split}
\end{equation}
Again, by the same argument of Lemma \ref{key lemma} we can show that $\hat{\bm a}^{\rm UL}$, the optimal BS association of the new problem $\mathrm P^{\mathrm{UL}}_{\mathrm{sum}} ( g_{nk}, p^{\mathrm{DL}}_n )$, is
still the unique solution to the assignment problem with gains $\{\log g_{ij}\}$.
Since Step 4 of \emph{DLSum} solves problem $(\mathrm P)$ with fixed association $\hat{\bm a}^{\rm UL}$ to global optima, we conclude that \emph{DLSumA} achieves $\gamma^*$. Q.E.D.

To finish the proof of Proposition \ref{prop: DLSum and optimal gamma}, suppose the min-SINR achieved by \emph{DLSum} is $\gamma^{\rm DL}$.
  If $\gamma^{\rm DL} \geq 1$, we infer that $\gamma^* \geq 1$ since the optimal min-SINR $\gamma^*$ must be no less than the achievable min-SINR.
  If $\gamma^* \geq 1$, according to Claim \ref{claim: DLSum globally solve the matching problem}, $\gamma^{\rm DL} = \gamma^* \geq 1$.
  Thus we have proved that $\gamma^{\rm DL} \geq 1$ if and only if $\gamma^* \geq 1$, i.e. \eqref{gamma relation 1}. Combining \eqref{gamma relation 1} and Claim \ref{claim: DLSum globally solve the matching problem}
  immediately leads to \eqref{gamma relation 2}.

\subsection{ Proof of Proposition \ref{propULconverge} }\label{appen: proof of UL_DL duality}

Denote
\begin{equation}\label{one power tight}
\bar{p}_{\mathrm{sum}} \triangleq \|\bar{\bm p} \|_1,
\end{equation}
then the power constraint of $(\rm P_{\rm sum}^{UL})$ becomes
\begin{equation}\label{one power tight}
\sum_k p_k \leq \bar{p}_{\mathrm{sum}}.
\end{equation}
\begin{lemma}\label{lemma of fixed point}
{\it
Suppose $(\bm p^*, \bm a^*)$ is an optimal solution to problem $(\rm P_{\rm sum}^{UL})$ (i.e. \eqref{max min, sum power, UL}), then $\bm p^*$ satisfies the following equation:
\begin{equation}\label{fixed point of max min}
\bm p^* = \frac{T(\bm p^*)}{\| \bm T(\bm p^*) \|_{1}  } \bar{p}_{\mathrm{sum}},
\end{equation}
or equivalently,
\begin{equation}\label{fixed point of max min, scaled}
\bm p^* = \frac{ T(\bm p^*)}{\|T(\bm p^*) \|_{\mathrm{s}} },
\end{equation}
where the scaled $\ell_1$ norm $\| \cdot \|_{\mathrm{s}}$ is defined as
$$
\| \bm x \|_s = \frac{\|\bm x \|_1}{ \bar{p}_{\rm sum} }.
$$

}
\end{lemma}
\noindent\emph{Proof of Lemma \ref{lemma of fixed point}}:
For a given power allocation $\bm p^*$, the optimal BS association is $a_k^* =  A_k (\bm p^*) = \arg \min_{ n } T_k^{(n)}(\bm p^*).$
Therefore, the SINR of user $k$ at optimality is
\begin{equation}\label{expression of SINR_k*}
 \text{SINR}_k^* = \frac{p_k^* }{ T_k^{(a^*_k)}(\bm p^*) } =\frac{ p_k^* }{  \min_{ n } T_k^{(n)}(\bm p^*) } = \frac{ p_k^*  }{ T_k(\bm p^*)}.
\end{equation}

Let $ \gamma^* $ denote the optimal value $ \min_k \text{SINR}_k^* $, then we have
\begin{equation}\label{SINR_k are equal}
\text{SINR}_k^* = \gamma^*,\quad \forall\; k.
\end{equation}
In fact, if $\text{SINR}_j^* > \gamma^*$ for some $j$, then we can reduce the power of user $j$ so that $\text{SINR}_j$ decreases and all other $\text{SINR}_k$ increase, yielding a minimum SINR that is higher than $\gamma^*$. This contradicts the optimality of $\gamma^*$, thus (\ref{SINR_k are equal}) is proved.

According to (\ref{expression of SINR_k*}) and (\ref{SINR_k are equal}), we have
\begin{equation}\label{T over p is gamma^*}
 \gamma^* T_k(\bm p^*)   = p_k^*, \quad \forall\; k.
\end{equation}

Next, we show that the maximum sum power is achieved at optimality, i.e.
\begin{equation}\label{one power tight}
\sum_k  p_k^* = \bar{p}_{\mathrm{sum}}.
\end{equation}
Assume $ \mu =  \frac{ \sum_k  p_k^*  }{ \bar{p}_{\mathrm{sum}} } < 1 $. Define a new power vector $ \bm p = \bm p^*/\mu$, then $\bm p$ satisfies the power constraints $\sum_k  p_k^* = \bar{p}_{\mathrm{sum}}.$ The SINR of user $k$ achieved by $(\bm p, \bm a^*)$ is
 \begin{equation}\label{beter SINR}
 \text{SINR}_k = \frac{p_k}{ T_k^{ (a^*_k) }(\bm p) }  =  \frac{p^*_k }{  \mu T_k^{ (a^*_k) }(\bm p^* /\mu)  } > \frac{ p^*_k }{ T_k^{ (a^*_k) }(\bm p^*) } = \text{SINR}_k^* ,
 \end{equation}
 which contradicts the optimality of $(\bm p^*, \bm a^*)$; therefore, \eqref{one power tight} is proved. The inequality in \eqref{beter SINR} is due to the fact that for any $k,n$, $\bm p $ and $0< \mu <1 $,
 $$
  \mu T_k^{ (n) }(\bm p /\mu) = \mu \frac{   \delta_{n}^2 + \sum_{j \neq k} g_{n j} p_j/ \mu }{ g_{n k} }
  = \frac{    \mu \delta_{n}^2 + \sum_{j \neq k} g_{n j} p_j }{ g_{n k} }  < \frac{  \delta_{n}^2 + \sum_{j \neq k} g_{n j} p_j }{ g_{n k} }  =T_k^{ (n) }(\bm p).
 $$

Plugging (\ref{T over p is gamma^*}) into (\ref{one power tight}), we obtain
\begin{equation}\label{gamma* expressed by norm of T}
\sum_k  \gamma^* T_k(\bm p^*) = \bar{p}_{\mathrm{sum}} \Longrightarrow  \gamma^* = \frac{ \bar{p}_{\mathrm{sum}} }{  \sum_k T_k(\bm p^*)  } = \frac{ \bar{p}_{\mathrm{sum}} }{ \| T(\bm p^*)\|_1  } .
\end{equation}
Plugging (\ref{gamma* expressed by norm of T}) into (\ref{T over p is gamma^*}), we obtain
$$
p_k^* = \frac{ \bar{p}_{\mathrm{sum}} }{ \| T(\bm p^*)\|_1  } T_k(\bm p^*),
$$
i.e. (\ref{fixed point of max min}). Q.E.D.

By definition \eqref{def of T_k}, the mapping
 $T(\bm p)=(T_1(\bm p), \dots, T_K(\bm p)): \mathbb{R}_{+}^K \rightarrow \mathbb{R}_{+}^K$ is the pointwise minimum of affine linear mappings $T^{(n)}(\bm p ) = (T_1^{(n)}(\bm p ), \dots, T_K^{(n)}(\bm p) )$, for $n=1,\dots, N$. It follows that $T$ is a concave mapping (i.e. each component function of $T$ is a concave function).
According to the concave Perron-Frobenius theory \cite[Theorem 1]{krause2001concave}, (\ref{fixed point of max min, scaled}) has a unique fixed point
, denoted as  $\bm p^{\rm UL}$, and the sequence $\{\bm p(t) \}$ generated by \emph{ULSum} converges to $\bm p^{\rm UL}$.
According to Lemma~\ref{lemma of fixed point}, any optimal power vector of problem $(\rm P_{\rm sum}^{UL})$ is a fixed point of (\ref{fixed point of max min}), thus problem $(\rm P_{\rm sum}^{UL})$ has a unique optimal power vector $\bm p^{\rm UL}$ and $\{ \bm p(t) \}$ converges to $\bm p^{\rm UL}$. 

To show the geometric convergence, we define
\begin{equation}\label{unit ball surface}
  U \triangleq \{\bm p \mid \| \bm p\|_{\mathrm{s}} = 1 \} = \{ \bm p \mid \|\bm p \|_{1} = \bar{p}_{\mathrm{sum}} \}.
\end{equation}
Then we have
$$
 \frac{  \delta_{n}^2 }{ g_{n k} }  \leq  T_k^{(n)}(\bm p) = \frac{  \delta_{n}^2 + \sum_{j \neq k} g_{n j} p_j }{ g_{n k} }
 \leq \frac{  \delta_{n}^2 + \bar{p}_{\mathrm{sum}} \max_{j } g_{n j}  }{ g_{n k} } , \forall \; \bm p \in U.
$$
Therefore,
 \begin{equation}\label{lower upper}
  A_k \leq T_k(\bm p) \leq  B_k , \quad \forall\; \bm p \in U ,
  \end{equation}
  where
  \begin{equation}
   A_k \triangleq \min_n \frac{\delta_n^2  }{g_{nk}} >0, \quad       B_k  \triangleq \min_n \frac{  \delta_{n}^2 + \bar{p}_{\mathrm{sum}} \max_{j } g_{n j}  }{ g_{n k} } .
       \end{equation}
  are both constants that only depend on the problem data.
  For two vectors $x, y$, we denote $x \geq y $ if $x_k \geq y_k,\ \forall\; k$.
   Define
   \begin{equation}
    \kappa = 1-\min_k \frac{A_k}{B_k} \in (0,1)
    \end{equation}
      and $$ \bm e = (B_1,\dots, B_K) >0 .$$
  Then (\ref{lower upper}) implies
 \begin{equation}\label{boundness of T}
  (1 - \kappa) \bm e \leq T(\bm p) \leq \bm e , \quad \forall\; \bm p \in U .
 \end{equation}
 According to the concave Perron-Frobenius theory  \cite[Lemma 3, Theorem]{krause1994relative},
if $T$ is a concave mapping and satisfies (\ref{boundness of T}), then the fixed point algorithm \emph{ULSum} converges geometrically at the rate $\kappa$.

We then prove the convergence to the optimal BS association set $\mathcal{A}^{\rm UL}$ by contradiction.
Assume the contrary, that there exists an infinite sequence $a(t_i), i=1,2,\dots$ such that
\begin{equation}\label{at def}
 a(t_i) \notin \mathcal{A}^{\rm UL} .
 \end{equation}
Define $f_k(\bm p, \bm a)$ as the function that maps a power vector $\bm p$ and a BS association $\bm a$ into the corresponding SINR of user $k$, $k=1,2,\dots,K$. Note that $f_k$ is a continuous function of $\bm p$.
The algorithm \emph{ULSum} generates a sequence $(\bm p(t), \bm a(t))$ that $f_k(\bm p(t), \bm a(t)) \longrightarrow \gamma^*, k=1,\dots, K$.
Since $\bm p(t) \longrightarrow \bm p^{\rm UL}$ and $f_k$ is continuous over $\bm p$, we have
\begin{equation}\label{converge of f}
f_k(\bm p^{\mathrm{UL}}, \bm a(t) ) \longrightarrow \gamma^*, \; \; t \longrightarrow \infty.
\end{equation}

Note that $\mathcal{A}^{\rm UL} = \{\bm a \mid f_k(\bm p^{\rm UL}, \bm a) = \gamma^* \}$.
Denote $(\mathcal{A}^{\rm UL})^c$ as the complement of $\mathcal{A}^{\rm UL}$, i.e. $(\mathcal{A}^{\rm UL})^c$  is the set of BS associations that are not in $\mathcal{A}^{\rm UL} $.
Each $\bm a \in (\mathcal{A}^{\rm UL})^c$ corresponds to an SINR value $ f_k(\bm p^{\rm UL}, \bm a) $ that is strictly less than $\gamma^*$,
and there are a finite number of BS associations in the set $ (\mathcal{A}^{\rm UL})^c$.
Thus all these SINR values are bounded above by a constant that is strictly less than $\gamma^*$, i.e.
\begin{equation}\label{upper bound of Ac}
\gamma^c \triangleq   \max_{\bm a \in (\mathcal{A}^{\rm UL})^c  } f_k(\bm p^{\rm UL}, \bm a) < \gamma^*.
\end{equation}
According to \eqref{at def} and \eqref{upper bound of Ac}, we have $f_k(\bm p^{\rm UL}, \bm a(t_i)) \leq \gamma^c <  \gamma^* , \forall i$, which contradicts \eqref{converge of f}.



\section{REFERENCES}
\bibliographystyle{IEEEbib}
\vspace{0.3cm}

{\footnotesize
\bibliography{refs}
}

\end{document}